\documentclass[11pt,a4paper]{article}
\pdfoutput=1
\usepackage{jheppub}

\usepackage{multirow, graphicx,amssymb,url,mathrsfs,amsmath}
\usepackage{wrapfig,boxedminipage,setspace,subfigure,epsfig}
\usepackage{amsxtra,amstext,latexsym,dsfont,amsfonts}
\usepackage{color}
\usepackage[dvipsnames]{xcolor}
\usepackage{float}
\usepackage{slashed}

\DeclareMathOperator{\sech}{sech}

\newcommand{\dd}{\mathrm{d}}

\newcommand{\grad}{\nabla}

\newcommand{\ie}{{\it i.e.}}

\newcommand{\tQ}{{\tilde{Q}}}

\newcommand{\kyr}[1]{{\color{black}{#1}}}

\title{Linear-$T$ resistivity from low to high temperature: axion-dilaton theories}

\author[a]{Yongjun Ahn,}
\author[a]{Hyun-Sik Jeong,}
\author[b]{Dujin Ahn,}
\author[a]{Keun-Young Kim}

\emailAdd{yongjunahn619@gmail.com}
\emailAdd{hyunsik@gm.gist.ac.kr}
\emailAdd{dujinahn@gmail.com}
\emailAdd{fortoe@gist.ac.kr}

\affiliation[a]{School of Physics and Chemistry, Gwangju Institute of Science and Technology,
	Gwangju 61005, Korea}
\affiliation[b]{Department of Mathematics, Yonsei University, Seoul 03722, Korea}				

\abstract{The linear-$T$ resistivity is one of the hallmarks of various strange metals regardless of their microscopic details. Towards understanding this universal property, the holographic method or gauge/gravity duality has made much progress.
Most holographic models have focused on the low temperature limit, where the linear-$T$ resistivity has been explained by the infrared geometry. We extend this analysis to high temperature and identify the conditions for a robust linear-$T$ resistivity up to high temperature. This extension is important because, in experiment, the linear-$T$ resistivity is observed in a large range of temperatures, up to room temperature. In the axion-dilaton theories we find that, to have a robust linear-$T$ resistivity, the strong momentum relaxation is a necessary condition, which agrees with the previous result for the Guber-Rocha model. However, it is not sufficient in the sense that, among large range of parameters giving a linear-$T$ resistivity in low temperature limit, only very limited parameters can support the linear-$T$ resistivity up to high temperature even in strong momentum relaxation. We also show that the incoherent term in the general holographic conductivity formula or the coupling between the dilaton and Maxwell term is responsible for a robust linear-$T$ resistivity up to high temperature. 
}

\begin{document}
	\maketitle
	\section{Introduction} \label{intro}

It has been shown that strongly correlated electron systems may have interesting universalities regardless of microscopic details.  
For example, the resistivity ($\rho$) is linear in temperature ($T$)~\cite{Hartnoll:2016apf} 
\begin{equation}
\rho \sim T \,,
\end{equation}
in various strange metals such as cuprates, heavy fermions and pnictides with a remarkable degree of universality.\footnote{As other examples of universal properties, there are  the Hall angle \cite{Blake:2014yla,Zhou:2015dha,Kim:2015wba,Chen:2017gsl,Blauvelt:2017koq,Kim:2010zq} at finite magnetic field and Homes's law in superconductors~\cite{Homes:2004wv, Zaanen:2004aa, Erdmenger:2015qqa,Kim:2015dna,Kim:2016hzi,Kim:2016jjk}.}
This is a distinctive property compared with the ordinary metal case, in which $\rho \sim T^2$, a different universal property explained by the Fermi liquid theory.

%
%
%
However, this well known ``linear-$T$ resistivity'' puzzle has not been completely resolved because of the theoretical difficulty in dealing with strong correlation. As a novel and effective tool to address strong correlation in general, the holographic methods or gauge/gravity duality~\cite{Zaanen:2015oix, Ammon:2015wua,Hartnoll:2016apf} has been widely used. The basic idea is to understand strongly correlated systems by mapping them to the dual classical gravity systems. 
%
%
%
%

There have been many researches to understand the linear-$T$ resistivity by holographic methods. Most researches,  for example \cite{Charmousis:2010zz, Davison:2013txa, Gouteraux:2014hca, Kim:2015wba, Zhou:2015dha,  Ge:2016lyn, Cremonini:2016avj, Chen:2017gsl, Blauvelt:2017koq, Ahn:2017kvc, Salvio:2013jia,  Donos:2014cya, Kim:2014bza,Kim:2015sma},  have focused on the relation between the linear-$T$ resistivity and the  infrared (IR) geometry, which can be supported by matter fields and couplings. The IR geometry is characterized by three critical exponent: the dynamical critical exponent ($z$), hyperscaling violating exponent ($\theta$) and
charge anomalous parameter ($\zeta$). 
{In particular, Gout\'{e}raux~\cite{Gouteraux:2014hca}, for the first time, systematically studied these geometries with momentum relaxation and characterized their scalings in terms of $z, \theta$ and $\zeta$ as well as derived the temperature scaling of the resistivity }
\begin{equation} \label{rho1}
\rho \sim T^{x(z,\theta,\zeta)} 
\,,
\end{equation}
in {\it low} temperature limit. Here, $x$ is some function of critical exponents. It is an interesting result since it gives an understanding of the linear-$T$ resistivity based on the scaling properties of condensed matter systems, which are nicely geometrized in dual gravity models.


However, at the same time, it has an important limitation. The result \eqref{rho1} is valid only in {\it low} temperature limit in the sense that $T$ has to be very low compared with any other scales in given models. For example, with a chemical potential ($\mu$) and momentum relaxation (of which strength is denoted by $k$), $T/\mu \ll 1$ and $T/k \ll 1$ etc.  Indeed, such a condition enables us to compute the power $x(z,\theta,\zeta)$ analytically. 
However, noting that the linear-$T$ resistivity has been observed in a `large' range of temperatures, up to room temperature $\sim 300 K$ in experiments, the condition of  `low' temperature limit for \eqref{rho1} will be too restrictive.

To deal with this problem, we first have to quantify `low' or `high' temperature compared with `what'. We may choose the chemical potential $\mu$ as our reference, $T/\mu$. However, it is not clear the relation between `$\mu$' in holography and the chemical potential, \kyr{$\mu$}, in real world. 
\kyr{Most holographic models including ours belong to bottom-up models. Without a top-down construction with a precise field theory dual, the meaning of the `chemical potential' (`$\mu$')  is ambiguous. It is just a `chemical potential' (`$\mu$')  for `some' conserved $U(1)$ charge. Furthermore, even in the case they are exactly the same quantity, it is still possible that there is some difference in numerical values, for example,  $\mu \sim 10`\mu$' or $\mu \sim 0.01`\mu$'.}
Thus, for a reference scale, it will be better to choose an intrinsic scale in the model.  For example, 
because experimental results show that the resistivity is linear in $T$ up to  $T > T_c$ \kyr{from zero $T$}, where $T_c$ is the superconducting transition (critical) temperature, $T_c$ can play a role as a reference scale.  In other words, the `low' or `high' temperature can be defined by the temperature below $T_c$ or above $T_c$.\footnote{\kyr{One may think that $T_c$ is of order $\mu$ so there is no essential difference to use $T_c$ as a reference scale, compared with $\mu$.  This may be true qualitatively. However, it turns out that the value of the order one number is important. For example, see Fig. 3 of \cite{Jeong:2018tua}, where there are six cases. For all cases $T_c \sim \mu$ up to oder one number. However, only some of them exhibit the linear-$T$ resistivity above $T_c$ from zero $T$. }} 
\kyr{Of course, it is possible to have a different reference scale other than $T_c$ depending on the context. Our purpose here is to motivate why a simple notion of small temperature based on $T/\mu < 1$ might be misleading in some cases. For example, in the context of  superconductor, $T_c$ is a better reference scale than $\mu$ because we have to check if the linear-$T$ resistivity persists up to high temperature $T>T_c$. We refer to \cite{Jeong:2018tua} for more details. }

Ref.~\cite{Jeong:2018tua} addressed this issue for the first time, to our knowledge.  In this work, the Gubser-Rocha model~\cite{Gubser:2009qt} with momentum relaxation~\cite{Davison:2013txa, Zhou:2015qui, Kim:2017dgz} have been considered with a complex scalar field to trigger superconducting instability. 
It has been shown that these models can exhibit the linear-$T$ resistivity up to `high' temperature, i.e. above $T_c$ only if momentum relaxation is strong. The importance of strong momentum relaxation was first emphasized in \cite{Hartnoll:2014lpa}: it was argued that, if the momentum relaxation, which is extrinsic and non-universal, is strong (quick), transport can be governed by diffusion of energy and charge, which is intrinsic and universal.  Thus, the universality of the linear-$T$ resistivity emerges with strong momentum relaxation\footnote{The linear-$T$ resistivity may appear in weak momentum relaxation regime in the case of weakly-pinned charge density waves (CDWs), where the resistivity is governed by incoherent, diffusive processes which do not drag momentum and can be evaluated in the clean limit \cite{Delacretaz:2016ivq,  Amoretti:2017frz, Amoretti:2017axe}. See also \cite{Davison:2018ofp,Davison:2018nxm}. }.

The work in Ref.~\cite{Jeong:2018tua} is important because it is the work studying the linear-$T$ resistivity at `high' temperature, while most of holographic works~\cite{Davison:2013txa, Kim:2015wba, Gouteraux:2014hca, Zhou:2015dha, Ge:2016lyn, Charmousis:2010zz, Cremonini:2016avj, Chen:2017gsl, Blauvelt:2017koq, Ahn:2017kvc} have been focused in the low temperature limit. However, Ref.~\cite{Jeong:2018tua} deals with only one class of models based on the Gubser-Rocha model, so it is not clear if strong momentum relaxation is really necessary and/or sufficient to have the linear-$T$ resistivity in general. The goal of this paper is to investigate this issue in a more general setup.


We start with a most general scaling geometry studied in~\cite{Davison:2013txa, Gouteraux:2014hca, Zhou:2015dha, Ge:2016lyn, Cremonini:2016avj, Chen:2017gsl, Blauvelt:2017koq, Ahn:2017kvc}, so called axion-dilaton theories or the Einstein-Maxwell-Dilaton with Axion model (EMD-Axion model). The axion field is introduced to realize the momentum relaxation effect.  The dilaton field is introduced with some potentials and couplings charaterized by three parameters $(\alpha, \beta, \gamma)$, in order to support  
the IR geometry parametrized by three scaling exponents ($z, \theta, \zeta$) as explained above \eqref{rho1},\footnote{These three exponents are related with three action parameters ($\alpha, \beta, \gamma$).} which is  rich enough to explore various possibilities.  
The IR geometry with an emblackening factor is valid only at {\it low} temperature limit. For the arbitrary temperature solutions, we need to introduce potentials and couplings supporting asymptotically ultraviolate (UV) AdS geometry~\cite{Kiritsis:2015oxa, Ling:2016yxy, Bhattacharya:2014dea}.  In general, there are many possibilities for potentials and couplings, and in this paper we consider a minimal (one-parameter) UV completion for potentials without changing couplings. This UV completion was introduced in \cite{Ling2017} for the purpose of studying the shear viscosity to entropy ratio and includes the Gubser-Rocha model in \cite{Jeong:2018tua} as a special case.

In short, our strategy is i) start with the various IR geometry giving the linear-$T$ resistivity such that $x(z,\theta,\zeta)=1$ in \eqref{rho1}, which is valid only in low $T$ limit; ii) after UV completing the geometry, obtain the arbitrary finite $T$ dependence of resistivity; iii) change the momentum relaxation parameter to see how it affects the robustness of linear-$T$ resistivity at high temperature. As a result, we have found that, in general, the strong momentum relaxation is still necessary to have a robust linear-$T$ resistivity up to higher temperature, but not sufficient: the parameter range for the robust linear-$T$ resistivity is quite limited compared with the high possibility in the low temperature limit~\cite{Gouteraux:2014hca, Kim:2015wba, Zhou:2015dha, Charmousis:2010zz, Davison:2013txa, Ge:2016lyn, Cremonini:2016avj}. We have identified this parameter range which is different from the Gubser-Rocha model.  In addition, we have also clarified the term which is responsible for the linear-$T$ resistivity in axion-dilaton theories: it is the incoherent term\footnote{{The first term in \eqref{sigma} can be properly called `incoherent' which means `no momentum dragging' only for strong momentum relaxation, which is the very regime we are interested in. For weak momentum relaxation, there is an incoherent contribution from the second term too~\cite{Davison:2015bea}. The first term is sometimes called a pair-creation term, which is proper if there is no net charge. The first term also has an interpretation as the conductivity in the absence of heat flows~\cite{Donos:2014cya}.}} the first term in the conductivity formula \eqref{sigma} or the coupling between the Maxwell and dilaton fields (for spatial dimension 2).
		
		We organize the paper as follows : In section \ref{Set up}, we review axion-dilaton theories and its low $T$ limit properties focusing on the linear-$T$ resistivity. In section \ref{sec:3}, we classify all possible parameter range in the action to obtain the linear-$T$ resistivity in IR and explain our UV-completion.  In section \ref{sec:4}, we report our results showing a robust linear-$T$ resistivity up to high temperature and demonstrate the importance of strong momentum relaxation. 
In section \ref{sec:5}, we interpret our results in more detail. We identify the term responsible for the linear-$T$ resistivity. We also describe the effects of the parameters introduced in axion-dilaton theories on the temperature dependence of the resistivity. 
In section \ref{sec:conclusion} we conclude.
	\section{Axion-dilaton theories: a quick review}\label{Set up}
	
	\kyr{In this section, we make a very brief review on the axion-dilaton theory investigated in~\cite{Gouteraux:2014hca}.\footnote{See also \cite{Charmousis:2010zz} for the original work without axion.} The purpose of this section is to set the stage and collect the results that will be useful for our discussion later. We refer to the original work~\cite{Gouteraux:2014hca} or a review in \cite{Ahn:2017kvc} for more detailed explanation and derivations.}

	\subsection{Action and equations of motion}
			An action of generic EMD-Axion models (or axion-dilaton theories) can be expressed as follows,
		\begin{align}\label{action}
			&S = \int \dd t \dd^{d}x \dd r \sqrt{-g}\left( R + \mathcal{L}_m \right) \nonumber \,,\\[1ex]
			&\mathcal{L}_m = \displaystyle- \frac{1}{2}(\partial \phi)^2 - \frac{J(\phi)}{2}\sum_{i=1}^{d} (\partial \chi_i)^2 + V(\phi) - \frac{Z(\phi)}{4}F^{2} \,,
		\end{align}
		where  $\phi$ and $\chi_{i}$ are scalar fields which are called dilaton  and axion respectively. The terms denoted by $J$, $Z$, and $V$ are the coupling functions and potential function.

The action yields the following Einstein equations:
%
%
\begin{align} \label{eommaster}
\begin{split}
 R_{\mu\nu} &= T_{\mu\nu} - \frac{1}{d}g_{\mu\nu}T\\
&=\frac{1}{2}\partial_{\mu}\phi\partial_{\nu}\phi
+\frac{J(\phi)}{2}\sum_{i=1}^{d}\partial_{\mu}\chi_{i}\partial_{\nu}\chi_{i}+\frac{Z(\phi)}{2}F_{\mu}{^\rho}F_{\nu\rho} -\frac{Z(\phi)F^2}{4d}g_{\mu\nu}-\frac{V(\phi)}{d}g_{\mu\nu} \,, \\
\end{split}
\end{align}
where $T_{\mu\nu} := -\frac{1}{\sqrt{-g}}\frac{\delta(\sqrt{-g}\mathcal{L}_{m})}{\delta g^{\mu\nu}}$ and $T = g^{\mu\nu}T_{\mu\nu}$ \,. The Maxwell equation, scalar equation, and axion equation are
\begin{align} \label{eommaster2}
\begin{split}
&\grad_{\mu}(Z(\phi)F^{\mu\nu}) = 0 \,, \\& \square\phi+V'(\phi)-\frac{1}{4}Z'(\phi)F^2-\frac{1}{2}J'(\phi)\sum_{i=1}^{d}(\partial\chi_{i})^2 =0
  \,, \\
&\grad_{\mu}(J(\phi)\grad^{\mu}\chi_{i}) =0\,.
\end{split}
\end{align}

By considering the following homogeneous (meaning all functions are only functions of $r$) ansatz
\begin{align} \label{backmaster}
\begin{split}
&\dd s^2=-D(r)\dd t^2+B(r)\dd r^2+C(r)\sum_{i=1}^{d}\dd x_{i}^{2}\,,\\
&\phi=\phi(r) \,, \quad A=A_t(r) \dd t \,, \quad \chi_{i}=k x_{i}\,,
\end{split}
\end{align}
we obtain the Einstein equations 
 %
\begin{align}
&  \!  0=\frac{Z(d-1)A_{t}'^{2}}{d D}+\frac{2BV}{d}+\frac{B'D'}{2BD} -\frac{d D'C'}{2DC}+\frac{D'^{2}}{2D^{2}}-\frac{D''}{D}\,,  \label{eom11} \\
&  \!  0 =\phi'^{2}-\frac{d C'^{2}}{2C^{2}}-\frac{d C'D'}{2CD}-\frac{d C' B'}{2CB} + \frac{d C''}{C}\,, \label{eom12} \\
&  \! 0=\frac{J k^{2}B}{C}-\frac{2BV}{d}+\frac{ZA_{t}'^{2}}{d D}+\frac{C'}{2C}\left(\frac{D'}{D}-\frac{B'}{B}\right)+\frac{(d-2)C'^{2}}{2C^{2}}+\frac{C''}{C}\,, \label{eom13} 
\end{align}
which come from the equations corresponding to $R_{tt}, R_{rr}$, and $R_{xx}$ in \eqref{eommaster} respectively.
The prime $'$ denotes the derivative with respect to $r$. The Maxwell equation and scalar equation are reduced to
\begin{align}
&0=\left[Z\frac{C^{\frac{d}{2}}}{\sqrt{BD}}A_{t}'\right]'\,, \label{max123} \\
&0=-\frac{d J_{,\phi}k^{2} B}{2C}+\frac{Z_{,\phi}A_{t}'^{2}}{2D}+BV_{,\phi}-\frac{B'\phi'}{2B} +\left(\frac{dC'}{2C}\right)\phi' +\frac{D'\phi'}{2D}+\phi'' \,, \label{axax}
\end{align}
and the axion equations are satisfied trivially.

\subsection{IR analysis of the axion-dilaton theories} \label{sec22}

Our task is to find the functions $B(r),C(r),D(r),A_t(r)$, and $\phi(r)$ in the ansatz \eqref{backmaster} satisfying the equations of motion \eqref{eom11}-\eqref{axax} for a given the couplings and potential functions in $J, Z$ and $V$ in \eqref{action}. 

In order to have analytic scaling solutions in IR (Infrared: far from the AdS boundary), we give the following constrains to $J, Z$ and $V$ in IR:
\begin{equation}\label{IRpot}
 V(\phi) \sim V_0 e^{\alpha \phi}\,, \qquad J(\phi) \sim e^{\beta \phi}\,, \qquad Z(\phi) \sim e^{\gamma \phi}\,,
\end{equation}
where the constant parameters ($\alpha, \beta, \gamma, V_0$) are introduced. We consider the coefficient $V_0$ only for $V$ without loss of generality because the overall factors for $J$ and $Z$ can be absorbed into the field $\chi_i$ and $A_t$.
Plugging the ``IR scaling coupling'' \eqref{IRpot} into the equations of motion \eqref{eom11}-\eqref{axax} {with the following scaling solution ansatz,}
\begin{equation}\label{scalinggeo}
\begin{split}
&\dd s^{2} = r^{\frac{2\theta}{d}}\left( -\frac{\dd t^2}{r^{2z}}+\frac{L^2\dd r^2}{r^2}+\frac{\sum_{i=1}^{d} \dd x_i^2}{r^2}\right)\,, \\ 
& A_t = Qr^{\zeta - z}\,, \quad \phi = \kappa \ln{(r)}\,, \quad \chi_i = kx_i\,,
\end{split}
\end{equation}
we obtain the solution ($z$, $\theta$, $\zeta$, $L$, $Q$, $\kappa$) in terms of $(\alpha, \beta, \gamma, V_0, k)$. We will call the exponents ($z$, $\theta$, $\zeta$) and the coefficients ($L, Q ,\kappa$) ``solution parameters'' or ``output parameters''.  We will call the parameters $(\alpha, \beta, \gamma, V_0, k)$ ``action parameters'' or ``input parameters''\footnote{In fact, $k$ is not introduced in an ``action'' level but we will include it in ``action parameters'' for convenience because it is one of the ``input parameters.''}.  Physically, $Q$ is proportional to the charge density and $k$ means the strength of the momentum relaxation.
Note that $\phi \to \pm \infty$ in IR\footnote{The IR regime in general can be near $r \to 0$ or $r \to \infty$. Without loss of generality, we may choose $r \to \infty$ as IR.} as $r \to \infty$ where $J \sim r^{\kappa\beta}, Z \sim r^{\kappa\gamma}$ and $V \sim r^{\kappa\alpha}$. 

After plugging scaling solution ansatz \eqref{scalinggeo} into equations of motion \eqref{eom11}$\sim$\eqref{axax}, we find that there are four possibilities to satisfy the equations of motion.
We may classify these solutions according to the ``relevance'' of the axion and/or charge, following \cite{Gouteraux:2014hca}.  By ``marginally relevant axion'' we mean the axion parameter $k$ appear explicitly in the leading solutions and by ``marginally relevant charge'' we mean $Q$ appears explicitly in the leading solutions. By ``irrelevant axion (charge)'' 
we mean $k$ $(Q)$ do not appear explicitly in the leading solutions but they can appear in the sub-leading solutions. Therefore, we may  consider four classes as follows. 

\begin{itemize}
	\item{class I: marginally relevant axion $\&$ charge \qquad \qquad \qquad \ \ \ \ ($k \ne 0, Q \ne 0$) }
	\item{class II: marginally relevant axion $\&$ irrelevant charge \qquad \, ($k \ne 0, Q = 0$) }
	\item{class III: irrelevant axion $\&$ marginally relevant charge \qquad \,($k = 0, Q \ne 0$) }
	\item{class IV: irrelevant axion $\&$ charge \qquad \qquad \qquad \qquad \qquad  \ \ \ ($k = 0, Q = 0$) }
\end{itemize}

Notice that the classification is based on the property of the leading solutions. To have a more complete picture, we also should consider the deformation by the sub-leading solutions:
\begin{equation} \label{perturb1}
	\Phi_i \rightarrow \Phi_i+ \epsilon_i r^{\beta_i} + \cdots  \,,
\end{equation}
where $\Phi_i$ denotes every leading order solution collectively, $\epsilon_i$ is a small parameter {and $\beta_i$ denotes the exponent of the sub-leading order and $\beta_i<0$  $(\beta_i>0)$ when the IR is located at $r\rightarrow\infty$ $(r\rightarrow0)$.}
Therefore, $Q =0 $ in the leading solution does not mean zero density and $k = 0$ in the leading solution does not mean no momentum relaxation because these parameters can appear in the sub-leading solutions.
In particular, if the axion is relevant, we may expect the momentum relaxation affects IR physics more strongly than the irrelevant axion cases. 

We present the explicit solutions for every classes in the following. 

\paragraph{Class I: charge and  axions are marginally relevant}
The leading order solutions read
\begin{equation}\label{classIexp}
	\begin{split}
	&z = \frac{-2 + d\alpha^2 - d \beta^2}{d(\alpha-\beta)\beta}\,, \qquad \theta = \frac{d \alpha}{\beta}\,, \qquad \zeta = \frac{\alpha + \gamma}{\beta} \,, \qquad \kappa = -\frac{2}{\beta} \,,  \\
	&L^2 = -\frac{2(d-\theta+z-1)(d-\theta+z)}{(d-1)k^2 - 2V_0}\,, \qquad Q^2 = \frac{2(d z - \theta)k^2 - 4V_0(z-1)}{((d-1)k^2-2V_0)(d-\theta+z)}\,, 
	\end{split}
\end{equation}
as well as a constraint between input parameters:
\begin{equation} \label{con000}
 \gamma = (d-1)\alpha - d\beta \,.
\end{equation}
or output parameters:
\begin{equation}
d+\zeta - \theta = 0  \,.
\end{equation}

\paragraph{Class II: charge is irrelevant; axions are marginally relevant}
The leading order solutions read
\begin{equation}\label{classIIexp}
	\begin{split}
	&z = \frac{-2 + d \alpha^2 - d \beta^2}{d(\alpha - \beta)\beta}\,, \qquad \theta = \frac{d \alpha}{\beta}\,, \qquad \kappa = -\frac{2}{\beta} \,,   \\
	&L^2 = \frac{(d-\theta+z)(d z - \theta)}{V_0}\,, 
	\end{split}
\end{equation}
as well as a constraint between input parameters:
\begin{equation} \label{con2}
k^2 = \frac{2V_0(z-1)}{d z - \theta} \,.
\end{equation}
Note that the $Q$ and $\gamma$ does not appear in the solutions \eqref{classIIexp} and the constrain \eqref{con2} can be understood as the condition $Q=0$ in \eqref{classIexp} so $\zeta$ is also undetermined. 

The value of $Q$ and $\zeta$ and their $\gamma$ dependences will be determined in the subleading order
\begin{equation}\label{gaugemode}
\zeta = d - \kappa\gamma  - \frac{d-2}{d}\theta = d + \frac{2\gamma}{\beta} - \frac{(d-2)\alpha}{\beta} \,.
\end{equation}
As explained in \eqref{perturb1}, this subleading order solution will back-react to metric and dilaton field, which behave as  
\begin{equation}\label{gaugebackreact}
	\sim r^{d+\zeta-\theta}\,.
\end{equation}
To have a stable IR geometry, we impose the constraint
\begin{equation} \label{con0000}
d+\zeta - \theta < 0  \,,
\end{equation}
near IR, $r \rightarrow \infty$. 
In terms of $\alpha, \beta, \gamma$ this constraint means
\begin{equation}\label{c2const}
2d - \frac{2(d-1)\alpha}{\beta} + \frac{2\gamma}{\beta}<0\,.
\end{equation}

\paragraph{Class III: charge is marginally relevant; axions are irrelevant}

The leading order solutions read
\begin{equation}\label{classIIIexp}
	\begin{split}
	&z = \frac{4d-\kappa^2(d \alpha^2 -2)}{2d(2+\alpha \kappa)}\,, \quad \theta = \frac{d^2 \alpha}{(d-1)\alpha - \gamma}\,,\quad \zeta = -\frac{1}{2}\kappa(\alpha + \gamma) \,, \quad \kappa = -\frac{2d}{(d-1)\alpha - \gamma}\,,  \\
		&L^2 = \frac{(d-\theta+z-1)(d-\theta+z)}{V_0}\,, \qquad Q^2 = \frac{2(z-1)}{d-\theta+z}\,.
	\end{split}
\end{equation}
Note that the $k$ and $\beta$ does not appear in the solutions \eqref{classIIIexp} so, compared with class I and II solutions, $\kappa \beta = -2$ does not hold anymore. The solutions \eqref{classIIIexp} may be understood from \eqref{classIexp} by setting $k=0$ and replacing $\beta$ by using \eqref{con000}. 

Similarly to class II, in the subleading order, the axion $(\chi_i = k x_i)$ starts playing a role and its back-reaction to metric and dilaton field behaves as 
\begin{equation}\label{axionbackreact}
	\sim r^{2+\kappa \beta}\,.
\end{equation}
To have a stable IR, we impose the constraint
\begin{equation} \label{con000001}
2+\kappa \beta < 0  \,,
\end{equation}
in the IR, $r \rightarrow \infty$. 
In terms of $\alpha, \beta, \gamma$ this constraint means
\begin{equation}\label{constc3}
2+\frac{2d \beta}{\gamma - (d-1)\alpha}<0 \,.
\end{equation}

\paragraph{Class IV: charge and axions are irrelevant}
The leading order solutions read
\begin{equation}\label{classIVexp}
	z=1\,, \qquad \theta = \frac{d^2 \alpha^2}{d\alpha^2-2}\,, \qquad \kappa = -\frac{2d \alpha}{d \alpha^2 - 2} \,, \qquad 
	L^2 = \frac{(d-\theta)(d-\theta+1)}{V_0}\,.
\end{equation}
Note that the $Q$, $\gamma$, $k$, and $\beta$ do not appear in the solutions. The solutions \eqref{classIVexp} can be understood from \eqref{classIIIexp} by setting $Q=0$ or $z=1$. Similarly to case II the subleading gauge field yields
\begin{equation}
\zeta = d - \kappa \gamma - \frac{d-2}{d}\theta = \frac{2d(\alpha(\alpha+\gamma)-1)}{d \alpha^2 -2} \,.\, 
\end{equation}

Similarly to class II and III we have the following constraints, \eqref{gaugebackreact} and \eqref{axionbackreact}:
\begin{equation} \label{con00000}
d+\zeta - \theta < 0  \quad \mathrm{and} \quad 2+\kappa \beta < 0\,,
\end{equation}
or equivalently
\begin{equation}  \label{con00000yy}
\frac{2d(\alpha(\alpha+\gamma)-2)}{d \alpha^2-2}<0 \quad \mathrm{and} \quad 2 - \frac{2d\alpha \beta}{d \alpha^2 - 2} <0\,.
\end{equation}

\subsection{Resistivity in low temperature limit}

Furthermore, it turns out that the emblackening factor $f(r)$ 
\begin{equation} \label{emblack1}
f(r) =  1- \left(\frac{r}{r_H}\right)^{z+d-\theta} \,,
\end{equation}
can be turned on ($\dd t^2 \rightarrow f \dd t^2$ and $\dd r^2 \rightarrow \dd r^2/f$ in \eqref{scalinggeo}) for all classes. 
Then, the Hawking temperature from this emblackening factor allows us to study the axion-dilaton theories at low temperature. However, it should be emphasized that the solution with this emblackening factor is valid only for very low temperature compared with any other scales, for example, chemical potential and momentum relaxation. 

In this low temperature limit, the Hawking temperature $T$ and charge density $q$ can be expressed in terms of solution parameters $(z, \theta, \zeta, Q)$: \cite{Gouteraux:2014hca, Ahn:2017kvc}
\begin{equation}
	T := \frac{1}{4\pi} \left. \frac{|D'|}{\sqrt{DB}}\right|_{r_H} = \frac{|d+z-\theta|}{4\pi}r_H^{-z} \,, \label{HT123}
\end{equation}
\begin{equation}
	 q := \left.  \sqrt{\frac{C^d}{DB}}ZA_t' \right|_{r_H} = Q(\zeta-z)\,, \quad s := \left. 4\pi C^{d/2}\right|_{r_H} = 4\pi r_H^{\theta-d} \,, \label{Hq123}
\end{equation}
where the subscript H denotes horizon.
The transport coefficients, for instance electric conductivity or thermal conductivity, can be calculated~\cite{Gouteraux:2014hca, Donos:2014cya}. Especially the electric DC conductivity formula for the EMD-Axion model is given by
		\begin{align}
			\sigma_{DC} &= Z_H C_H^{\frac{d-2}{2}} + \frac{q^2}{k^2  C_H^{d/2} J_H} \label{sigma} \\
			 &\sim  T^{-(2-\zeta)/z} + \frac{q^2}{k^2}T^{-\frac{d-\theta-\kappa \beta}{z}}\,.
		\end{align}
%
Note that only the behavior of  ($Z, J, C$) at the horizon play a role in determining the power of $T$ via the relation between $r_H$ and $T$ \eqref{HT123}. More variables ($A_t, B, C, D, Z$) enter for the charge density $q$ as shown in \eqref{Hq123}.

By using the constraints between $\zeta, \beta,  \kappa$, \eqref{con000}, \eqref{con0000}, \eqref{con000001}, \eqref{con00000},
we can find which term in \eqref{sigma} is dominant in low temperature limit. In class I both two terms are of the same order, in class II the first term is dominant, in class III the second term is dominant, and in class IV the dominant term depends on the parameters. 
\ie 
$\,$ in low temperature limit the resistivity $\rho \sim \sigma_{DC}^{-1}$  behaves as:
\begin{equation}\label{eq:res}
\begin{split}
	&\text{Class I \ \ : } \rho \sim T^{(2+d-\theta)/z}\,,\\
	&\text{Class II \ : } \rho \sim T^{\frac{2-\zeta}{z}}\,,\\
	&\text{Class III : } \rho \sim T^{\frac{d-\theta-\kappa \beta}{z}}\,,\\
	&\text{Class IV : } \rho \sim T^{d-\theta-\kappa \beta} \quad \text{or} \, \, \, T^{2-\zeta}\,.
\end{split}
\end{equation}
Note that the power $x$ in $\rho \sim T^{x}$ does not depend on the momentum relaxation $k$. This is because this formula is valid only in low temperature limit.  At finite temperature, this is not the case as we will show.

In fact, in addition to the constrains, \eqref{con000}, \eqref{con0000}, \eqref{con000001}, \eqref{con00000}, there are more constraints\footnote{For more details, we refer to \cite{Gouteraux:2014hca,Ahn:2017kvc}. } for the exponents $z$ and $\theta$, which will be necessary to constrain our physical model later.  
First, since we are considering the case that IR is located at $r \rightarrow \infty$,\footnote{In principle, we may consider the case that IR is located at $r \rightarrow 0$. In this case we will end up with negative $z$ in \eqref{thetad}. We do not consider this case because it is not physical.} the exponent of $r$ of each metric component in \eqref{scalinggeo} should be negative, which implies
	\begin{equation}\label{IRcon}
		\theta < d z\,, \quad \theta < d\,.
	\end{equation}
Second, the emblackness factor \eqref{emblack1} $f(r) \rightarrow 1$ at the UV ($r \to 0$), which implies
	\begin{equation}\label{blackcon}
		\theta < z + d\,.
	\end{equation}
Third,  the condition that the specific heat is positive implies
	\begin{equation}
	\frac{d - \theta}{z} > 0 \,,
	\end{equation}
where we used the relation between the entropy and temperature \eqref{HT123}: $s \sim T^{\frac{(d - \theta)}{z}}$.
Fourth,  the $\kappa, Q,$ and $L$ in the \eqref{classIexp}, \eqref{classIIexp} and \eqref{classIIIexp} should be real. The $\kappa$ can be rewritten by :
\begin{equation}\label{kappa}
\kappa^2 = \frac{2(d-\theta)(d(z-1)-\theta)}{d}\,.
\end{equation}
The conditions \eqref{IRcon} - \eqref{blackcon} and $Q^2 > 0,\, L^2 > 0,\, \kappa^2>0$ imply
\begin{equation} \label{thetad}
z>1, \quad \theta < d \,.
\end{equation}

Note that Eqs. \eqref{con00000} and \eqref{thetad} imply that both $2-\zeta$ and $d-\theta- \kappa \beta$ can not be 1 in class IV.  Thus, the class IV solution does not allow the linear-$T$ resistivity in the low temperature limit. Being interested in the linear-$T$ resistivity, we study only class I, II and III solutions in the following section. 

\section{Resistivity up to high temperature}
\label{sec:3}

From \eqref{eq:res}, we may obtain the condition for the linear-$T$ resistivity. However, as we emphasized in the previous section, the temperature dependence of the resistivity \eqref{eq:res} is valid only in low temperature limit, which means that the temperature is very low compared with any other scales such as the chemical potential, the momentum relaxation strength, or the superconducting phase transition scale. However, from experimental results, it is important to have a robust linear-$T$ resistivity also at intermediate and high temperature, which we will call {\it finite} temperature to distinguish it from low temperature limit. 

\subsection{UV-completion of potentials}
At finite temperature, the couplings and potential in \eqref{IRpot} only valid in IR have to be UV-completed. In order to have asymptotically AdS space in UV, we impose the following conditions \kyr{\cite{Charmousis:2010zz}}: 
\begin{align} 
	&V(\phi) = \frac{(d+1)d}{\ell_{AdS}^2} - \frac{1}{2}m^2 \phi^2 + \cdots \,, \label{cond10} \\ 
	& Z(\phi) = 1 + \cdots  \,, \quad J(\phi) = 1 + \cdots \,, \label{cond11}
\end{align} 
near boundary ($r \sim 0$). In other words, for $V(\phi)$,
\begin{equation} \label{cond2}
	V(0) = -2\Lambda =\frac{(d+1)d}{\ell_{AdS}^2}\,, \qquad V'(0) = 0\,, \qquad V''(0) = -m^2 = \frac{-\Delta(\Delta - d - 1)}{\ell_{AdS}^2} \,,
\end{equation}
where $\Delta$ is the conformal dimension of the dual operator of $\phi$ and $\phi \rightarrow 0$ in UV.  For simplicity we set $\ell_{AdS} = 1$ from here.
%
%
%
%

In principle, there will be many possibilities to construct $V, Z, J$ satisfying \eqref{IRpot} in IR and \eqref{cond10} and \eqref{cond11} in UV. See for example~\cite{Gouteraux:2011ce, Kiritsis:2015oxa, Ling2017, Davison:2018nxm}. Here, for simplicity, we choose one minimal way studied in \cite{Ling2017}:
\begin{equation}\label{ZandJV1}
Z(\phi) = e^{\gamma \phi}\,, \qquad J(\phi) = e^{\beta \phi}\,,
\end{equation}
and three cases for $V(\phi)$
\begin{equation}\label{ZandJV2}
\begin{split}
		&V(\phi) = \left\{ \begin{array}{lll}
		\frac{2d}{\alpha^2}\sinh^2\left(\frac{\alpha \phi}{2}\right) + (d+1)d \,,                                                   \, & \text{for} \, \theta <0 \,,\\[0.5ex]
		(d+1)d\,,                                                                                                                                                                                  \,  & \text{for} \, \theta = 0 \,,\\[0,5ex]
		d\left(\frac{1}{\alpha^2}+2\left(d+1\right)\right) \sech(\alpha \phi)-d\left(\frac{1}{\alpha^2}+\left(d+1\right)\right)\sech^2\left(\alpha \phi \right), \, &\text{for} \, d>\theta>0 \,.
		\end{array} \right.
	\end{split}
\end{equation}

Let us now explain the rationale for our choice. All $V(\phi)$ in  \eqref{ZandJV2} satisfy the expansion \eqref{cond10}, where  
the dilaton mass $m^2 = -d$ for $\theta \ne 0$ and $m^2 = 0$ for $\theta =0$. The dilaton $\phi$ behaves near boundary as follows
	\begin{equation}
	\label{eq:dilaton}
	\phi(r) \sim \begin{cases}
			c + \cdots, \quad  \theta = 0\,,\\
			 r + \cdots, \quad \theta \neq 0\,,
			\end{cases}
	\end{equation}
where $c$ is a constant and we will set $c=0$. Thus, our choice of the dilaton mass makes \eqref{ZandJV1} satisfy the UV condition \eqref{cond11}. The particular choice of $V_0$ we make is not necessary but for simplicity. Thanks to this choice, our $V(\phi)$ becomes a function of only $\alpha$.\footnote{For example, if we do not choose the dilaton mass as $m^2=-d$ our potential will include $m^2$ as a free parameter.}

We consider three kinds of $V(\phi)$ depending on the sign of $\theta$ in \eqref{ZandJV2}. To understand the necessity of these three different forms let us first find the relation between the signs of $\alpha$ and $\theta$. 
Near IR $r \rightarrow \infty$, we choose $\phi \rightarrow +\infty$, which means $\kappa > 0$ from \eqref{scalinggeo}. When $\kappa>0$, the relations between $\alpha$ and $\theta$ in \eqref{classIexp}, \eqref{classIIexp}, \eqref{classIIIexp} imply that the signs of $\theta$ and $\alpha$ are opposite, and if $\theta=0$ then $\alpha=0$. Thus, we find that the asymptotic potentials near IR read
\begin{equation} \label{IRp123}
\begin{split}
		&V(\phi) = \left\{ \begin{array}{lll}
		\frac{d}{2\alpha^2}\left(e^{\alpha \phi} + e^{-\alpha \phi} - 2\right) + (d+1)d \approx V_0 e^{\alpha \phi}\,,                                                   & V_0 = \frac{d}{2\alpha^2}\,,                 & \text{for} \, \alpha >0\,,\\[0.5ex]
		(d+1)d\,,                                                                                                                                              & V_0 = (d+1)d\,,                                    & \text{for} \, \alpha = 0\,,\\[0,5ex]
		\frac{d\left(\frac{1}{\alpha^2}+2(d+1)\right)}{\cosh \left(\alpha \phi\right)} -\frac{d\left(\frac{1}{\alpha^2}+(d+1)\right)}{\cosh^2 \left(\alpha \phi\right)}\approx V_0 e^{\alpha \phi}\,,& V_0 = 2d\left(\frac{1}{\alpha^2}+2d +2\right), &\text{for} \, \alpha<0\,,
		\end{array} \right.
	\end{split}
\end{equation}
which precisely match to our IR condition $V \sim V_0 e^{\alpha\phi}$ in \eqref{IRpot}. For example, the first potential will take a form of $V_0 e^{-\alpha \phi}$ for $\alpha<0$, which is not consistent with our IR condition $V \sim V_0 e^{\alpha\phi}$ in \eqref{IRpot}.



Note that the first potential in \eqref{ZandJV2} includes the Gubser-Rocha model with linear axion. For $(\alpha, \beta, \gamma) = (1/\sqrt{3}, 0, -1/\sqrt{3})$ and $d = 2$,  $V(\phi), Z(\phi)$ and $J(\phi)$ in \eqref{ZandJV1} and \eqref{ZandJV2} yield
\begin{equation}\label{GRwithA}
V(\phi) = 6 \cosh{\frac{\phi}{\sqrt{3}}}\,, \qquad Z(\phi) = e^{-\frac{\phi}{\sqrt{3}}}\,, \qquad J(\phi) = 1 \,,
\end{equation}
which is nothing but the Gubser-Rocha model with linear axion studied in \cite{Gouteraux:2014hca, Jeong:2018tua}.

\subsection{Numerical methods} \label{sec32}
Because the potential is changing, the scaling solution  \eqref{scalinggeo} is not valid anymore so we start with the following ansatz :
	\begin{equation}\label{ansatz}
	\begin{array}{l}
	\dd s^{2} =\displaystyle \frac{1}{u^2}\left( - (1-u)U(u)e^{-S(u)} \dd t^2 + \frac{\dd u^2}{(1-u)U(u)} + \sum_{i=1}^{d}\dd x_i^2 \right)\,,\\
	\phi = \phi(u)\,, \qquad \chi_i = k x_i\,, \qquad A = (1-u)A_{t}(u) \dd t\,,
	\end{array}
	\end{equation}
	where $u := r/r_h$. In this coordinate, the horizon and the boundary are located at $u = 1$ and $u = 0$ respectively. Since we want the geometry to be asymptotically $AdS_{d+2}$ near boundary, we impose the conditions $U(0) = 1$ and  $S(0) = 0$. 	
	
	In our set-up, there are three dimensionful parameters: the chemical potential $\mu = A_t(0)$, the Hawking temperature $T$, and the momentum relaxation strength $k$.  For numerical analysis, we will take $d = 2$. We choose the chemical potential as our scale so the dimensionless parameters are $T/\mu$ and $k/\mu$. 
	
\begin{figure}[]
	\centering
	\includegraphics[width=0.4 \linewidth]{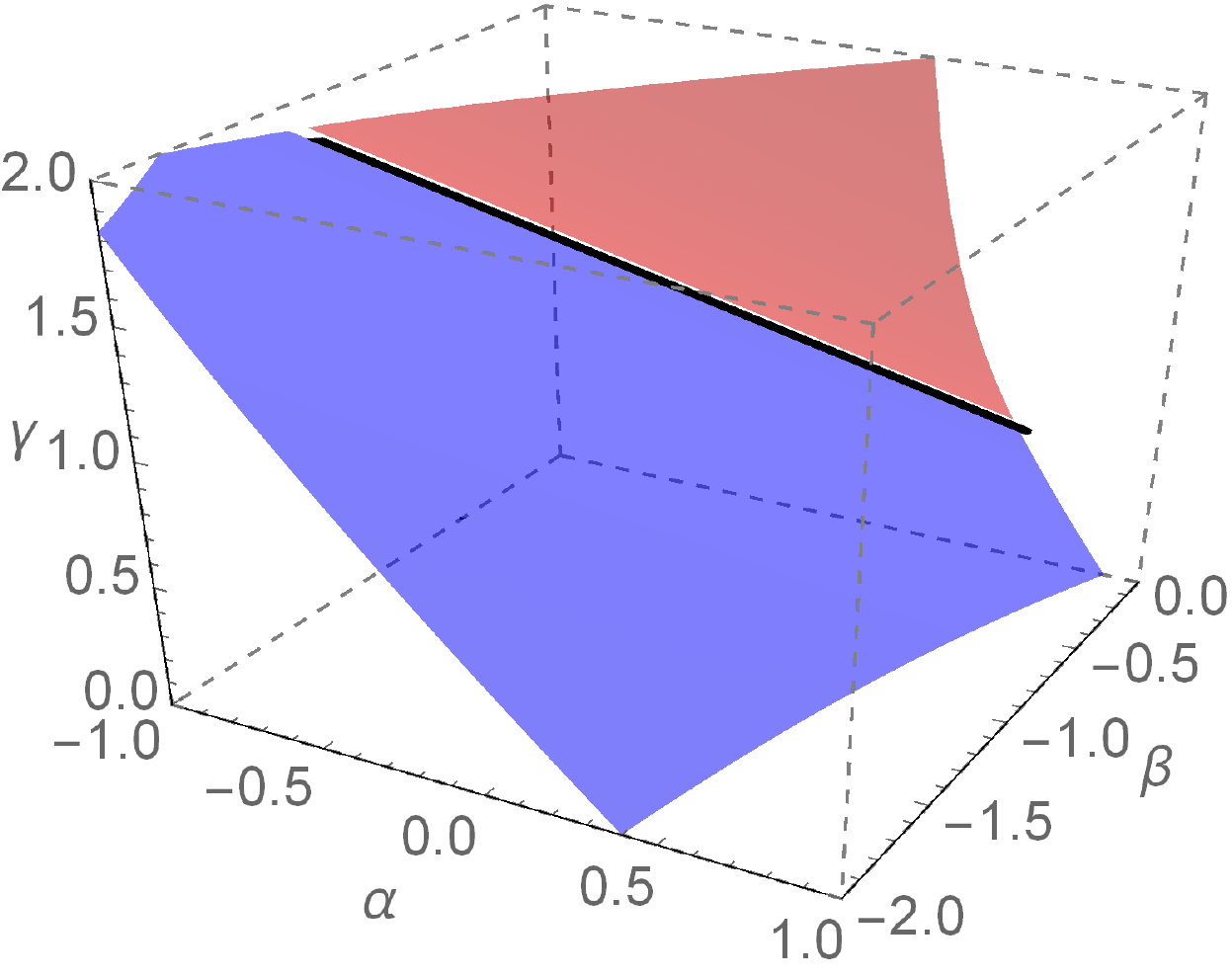}
	\caption{The $(\alpha, \beta, \gamma)$ region which gives the linear-$T$ resistivity in low temperature limit. The (red/blue) surface corresponds to the (class II/class III) solution. The black line, which is the intersection of the red and blue surface, corresponds to the class I. Class IV does not allow the linear-$T$ resistivity.}
	\label{fig:wregion}
\end{figure}

	Our controlling parameters are $(\alpha, \beta, \gamma)$, which determine the whole action as well as the IR behavior of the model via \eqref{ZandJV1} and \eqref{ZandJV2}. We want to find the parameter range $(\alpha, \beta, \gamma)$ which yield the linear-$T$ resistivity from low temperature to high temperature. Therefore, we start with the restricted range of $(\alpha, \beta, \gamma)$, which gives the linear-$T$ resistivity in low temperature regime. Fig.~\ref{fig:wregion} shows such a range in $(\alpha, \beta, \gamma)$ space.  It can be understood as follows.
\begin{itemize}
\item class I: The linear-$T$ condition \eqref{eq:res} ($(2+d-\theta)/z=1$ in terms of $\alpha, \beta, \gamma$ with \eqref{classIexp}) and \eqref{con000} gives a curve  in 3 dimensional $\alpha, \beta, \gamma$ space:
\begin{equation}
\left(\alpha, \alpha \pm \sqrt{\frac{2}{d(d+1)}}, -\alpha \mp \sqrt{\frac{2}{d(d+1)}}d \right) \,,
\end{equation}
which is the black line in Fig. \ref{fig:wregion}.
\item class II: The linear-$T$ condition \eqref{eq:res} ($(2-\zeta)/z = 1$ in terms of $\alpha, \beta, \gamma$ with \eqref{classIIexp}) defines a surface in 3 dimensional $\alpha, \beta, \gamma$ space:
\begin{equation}
\left(\alpha, \frac{d^2\alpha -d(2\alpha+\gamma)\pm\sqrt{d(2+d((\alpha+\gamma)^2-2))}}{d(d-1)}, \gamma \right).
\end{equation} 
However, the inequality \eqref{c2const} restricts the available surface, which is the red surface in Fig. \ref{fig:wregion}.
\item class III: The linear-$T$ condition \eqref{eq:res} (($d-\theta-\kappa \beta)/z=1$ in terms of $\alpha, \beta, \gamma$ with \eqref{classIIIexp}) defines a surface in 3 dimensional $\alpha, \beta, \gamma$ space: 
\begin{equation}
\left(\alpha, \frac{3\alpha^2+4\alpha \gamma + \gamma^2 -2 -\frac{(\alpha + \gamma)^2}{d}}{2(\alpha + \gamma)},\gamma \right).
\end{equation} 
However, the inequality \eqref{constc3} restricts the available surface, which is the blue surface in Fig. \ref{fig:wregion}.
\item class IV: There is no range. See the explanation below \eqref{thetad}.
\end{itemize}

The potential we consider is valid only for $\beta < 0$ because it corresponds to $\kappa > 0$ due to \eqref{classIexp}, \eqref{classIIexp} and \eqref{classIIIexp} so correctly match the IR potential \eqref{IRp123}.  If we want to consider the case $\beta>0$ we need to reconstruct the potential \eqref{ZandJV2} accordingly.  Note also that the sign of $\alpha$ is opposite to $\theta$ so i) for $\alpha > 0$ region, the first potential in \eqref{ZandJV2} should be used ii) for $\alpha = 0$, the second potential should be used. 
  iii) for $\alpha < 0$ region, the third potential in \eqref{ZandJV2} should be used.

\section{Linear-$T$ resistivity from low temperature to high temperature}  \label{sec:4}
	
We have fine-gridded the surface in Fig. \ref{fig:wregion}. By trying out all gridded data set ($\alpha, \beta, \gamma $), we have found that, only for a small range of parameters near the parameter set 
\begin{equation}
(\alpha, \beta, \gamma) = \left(-\frac{1}{\sqrt{3}}, -\frac{2}{\sqrt{3}}, \sqrt{3}\right) \ \ \Leftrightarrow \  \  (z, \theta, \zeta) = (3, 1, -1) \,,
\end{equation}
yields the linear-$T$ resistivity from low temperature to high temperature when the momentum relaxation is strong. 

\begin{figure}[]
\centering
     \subfigure[Resistivity ($\rho$) vs Temperature ($T/\mu$)]
     {\includegraphics[width=6.96cm, height = 4.7cm]{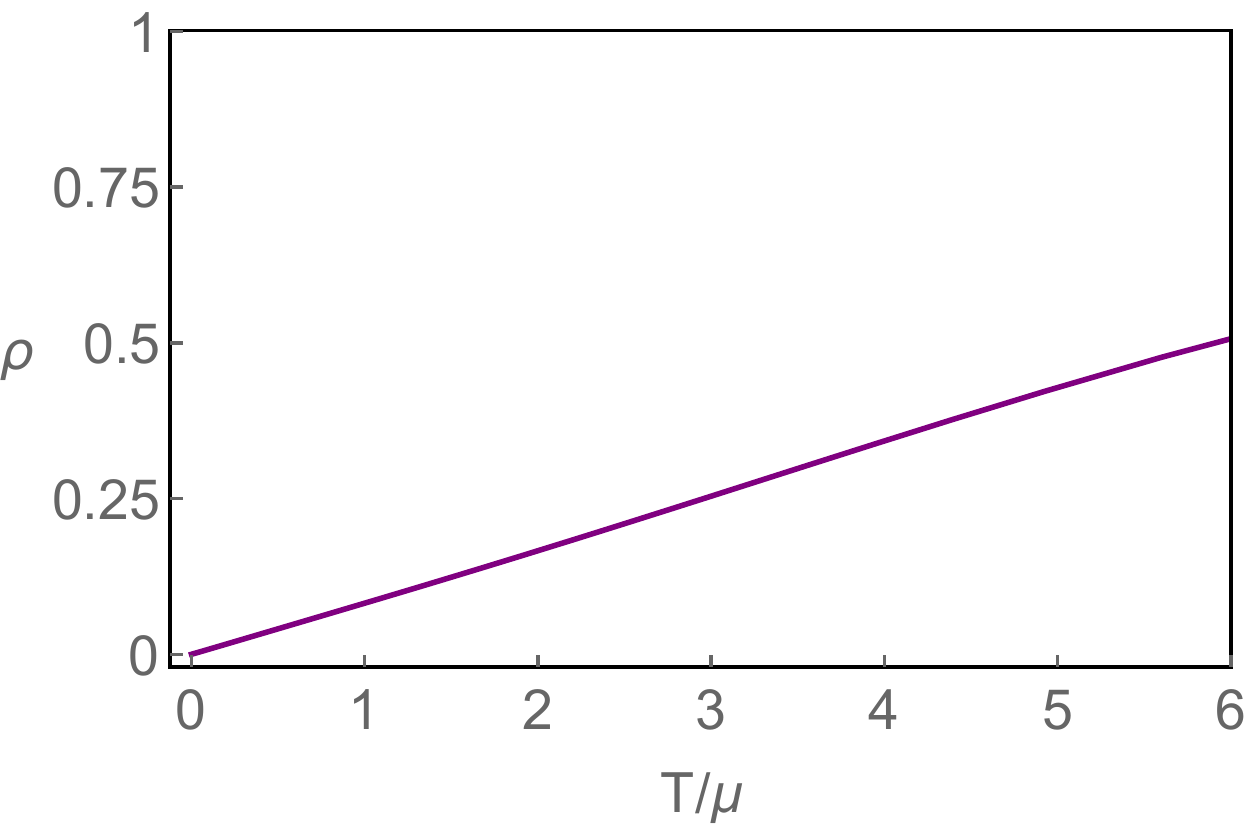} \label{fig:z3t1lrhoT}}\ \ \ 
      \subfigure[The power $x$ where $\rho \sim  T^x$. The red dashed line is a guide to the eye.]
     {\includegraphics[width=7.4cm, height = 4.7cm]{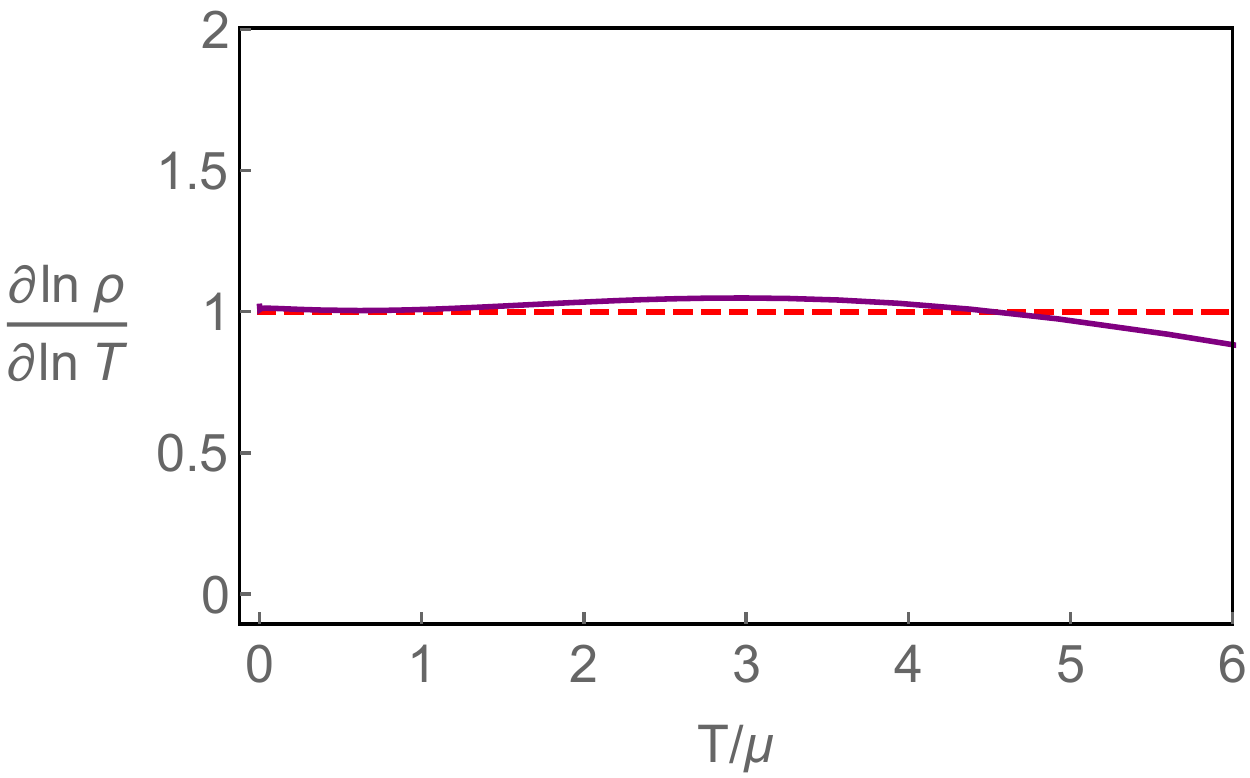} \label{fig:z3t1lrhoTp}}
 \caption{Resistivity vs temperature: the linear-$T$ resistivity.   ($\alpha, \beta, \gamma) = \left(-\frac{1}{\sqrt{3}}, -\frac{2}{\sqrt{3}}, \sqrt{3} \right)$  (i.e. $(z, \theta, \zeta) = (3, 1, -1)$) with large momentum relaxation $k/\mu = 20$. }\label{fig:z3t1}
\end{figure}

See Fig.~\ref{fig:z3t1lrhoT}  for the numerical results for this parameter set with a strong momentum relaxation $k/\mu = 20$. It is a plot for the resistivity ($\rho$) vs  temperature $T$, showing  $\rho \sim T^x$ with $x\sim 1$. 
\begin{figure}[]
\centering
     \subfigure[$(\alpha,\beta,\gamma) = \left(\frac{1}{3\sqrt{3}}, -\frac{2}{3\sqrt{3}}, \frac{5}{3\sqrt{3}}\right)$ (i.e.$\left(z, \theta, \zeta\right) = (5, -1, -3))$) with $k/\mu=10$. The black dot case in Fig.~\ref{fig:nthregion2}.]
     {\includegraphics[width=6.96cm, height = 4.7cm]{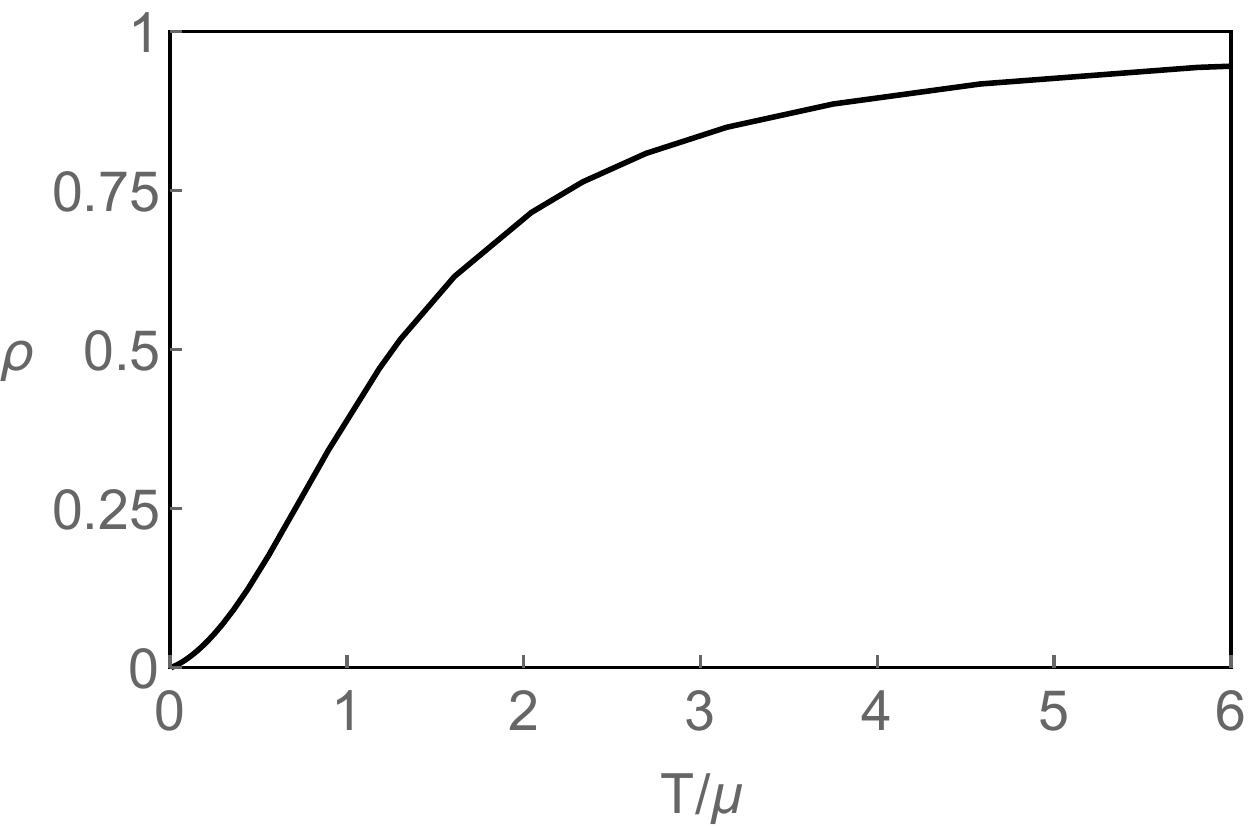} \label{new123}}\ \ \ \ \ \ 
     \subfigure[$(\alpha,\beta,\gamma) = \left(0,-\frac{1}{\sqrt{3}},\frac{2}{\sqrt{3}}\right)$ (i.e.$\left(z, \theta, \zeta\right) = (4 ,0 ,-2 )$) with $k/\mu=10$. The black dot case in Fig.~\ref{fig:zthregion}.]
     {\includegraphics[width=6.96cm, height = 4.7cm]{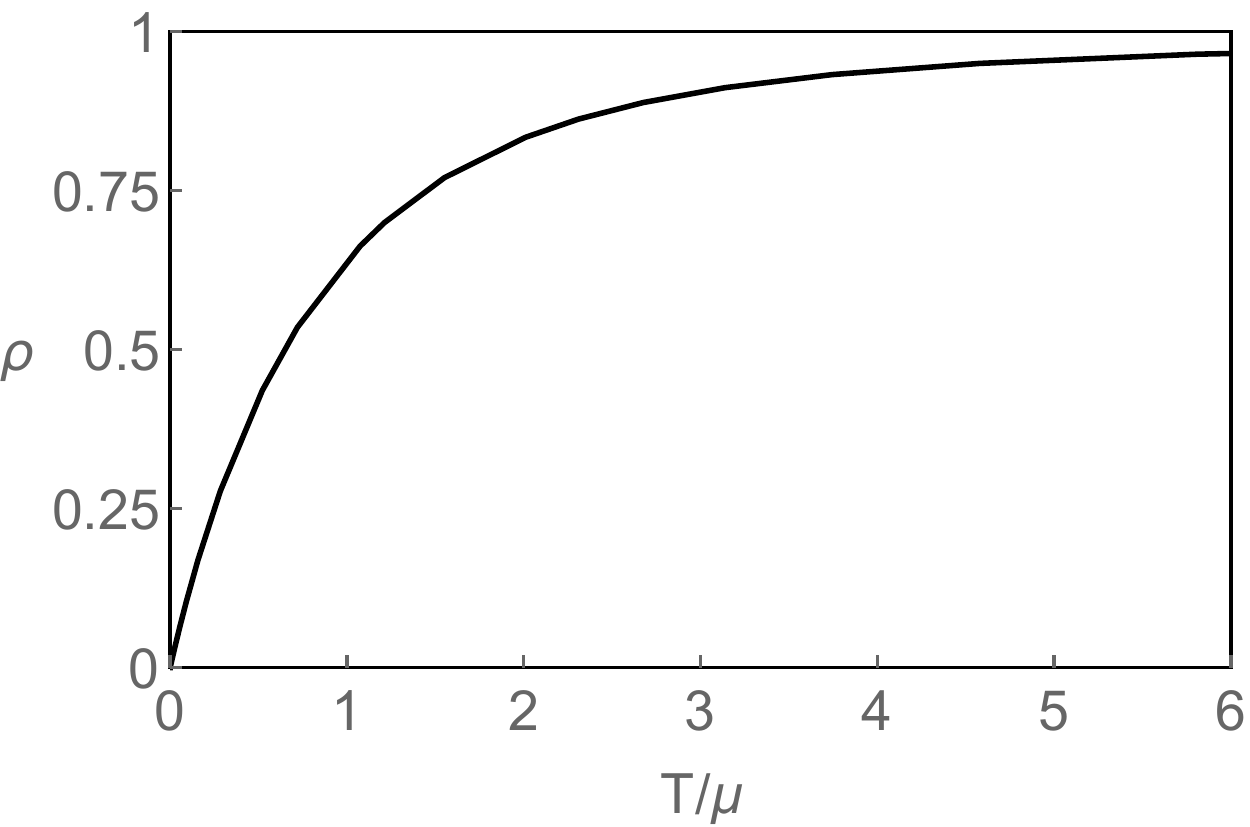} \label{new1234}}
 \caption{Resistivity vs temperature: non-linear-$T$ resistivity}\label{fig:z3t123}
\end{figure}
To show the value of the exponent $x$ more clearly, we  make another plot for $\frac{\partial \ln \rho}{\partial \ln T }$ in Fig.~\ref{fig:z3t1lrhoTp}, where we see $x\sim 1$ for $T/\mu \lesssim 5$.
As we explain in section \ref{sec43},  a small neighborhood of the point  $(-1/\sqrt{3}, -2/\sqrt{3}, \sqrt{3})$ also exhibits the linear-$T$ resistivity. 

For a purpose of comparison, we also show typical plots for non-linear-$T$ resistivity in Fig.~\ref{fig:z3t123}. The parameters used in Fig.~\ref{fig:z3t123} were chosen as the same as the one in Fig.~\ref{fig:nthregion2} and Fig.~\ref{fig:zthregion}. 
Note that  Fig.~\ref{fig:z3t1} corresponds to the third potential ($\alpha<0$) in 
\eqref{ZandJV2} while Fig.~\ref{new123} and Fig.~\ref{new1234} correspond to the first ($\alpha>0$) and second ($\alpha=0$) potential in \eqref{ZandJV2} respectively. It turns out that the third potential or $\alpha<0$ is more advantageous to have a linear-$T$ resistivity than the others.  We will discuss about it more in sec.~\ref{sec43}.


	\subsection{Momentum relaxation effect}\label{sec:z3t1zm1}
\begin{figure}[]
\centering
     \subfigure[$\rho$ vs $T/\mu$]
     {\includegraphics[width=6.96cm, height = 4.7cm]{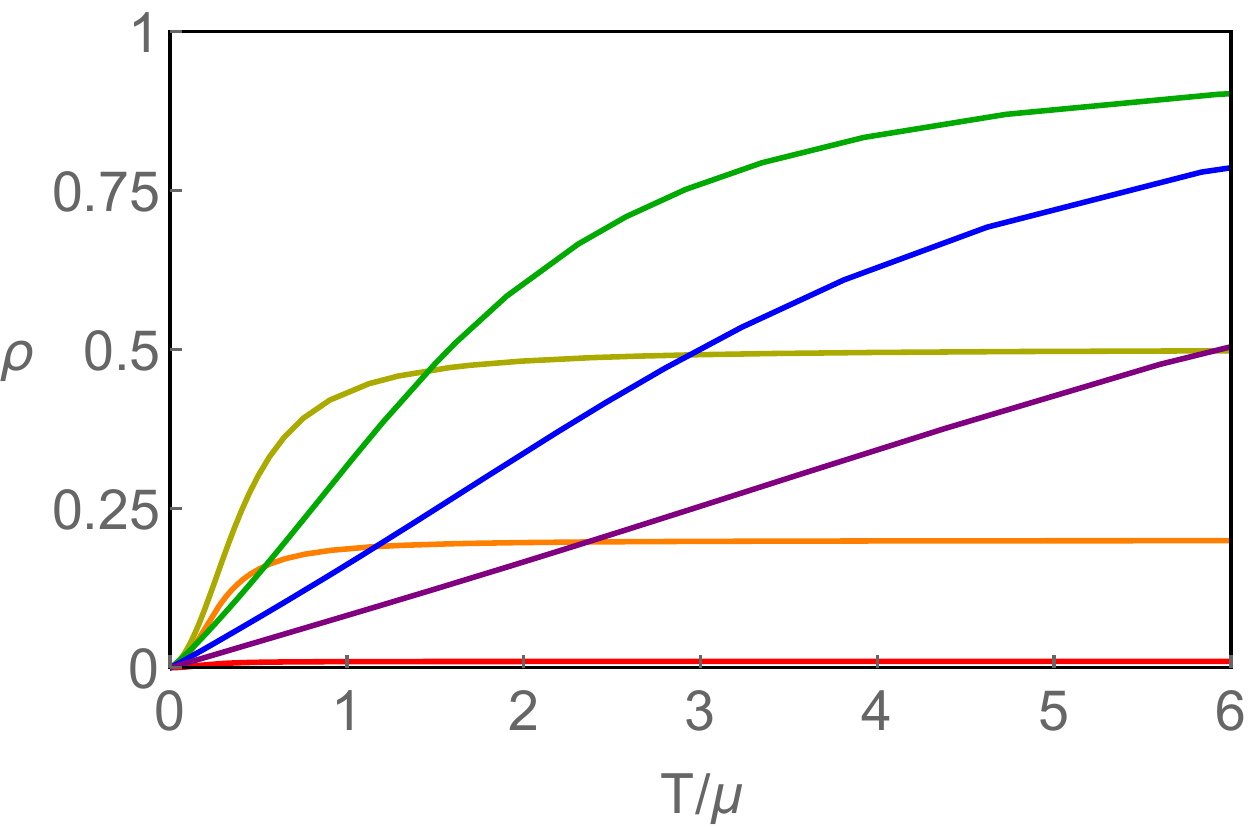} \label{fig:z3t1lrhoTvm}}\ \ \ 
     \subfigure[$x$ vs $T/\mu$ ($\rho \sim T^x$). The dotted line is for $x=1$.]
     {\includegraphics[width=7.4cm, height = 4.7cm]{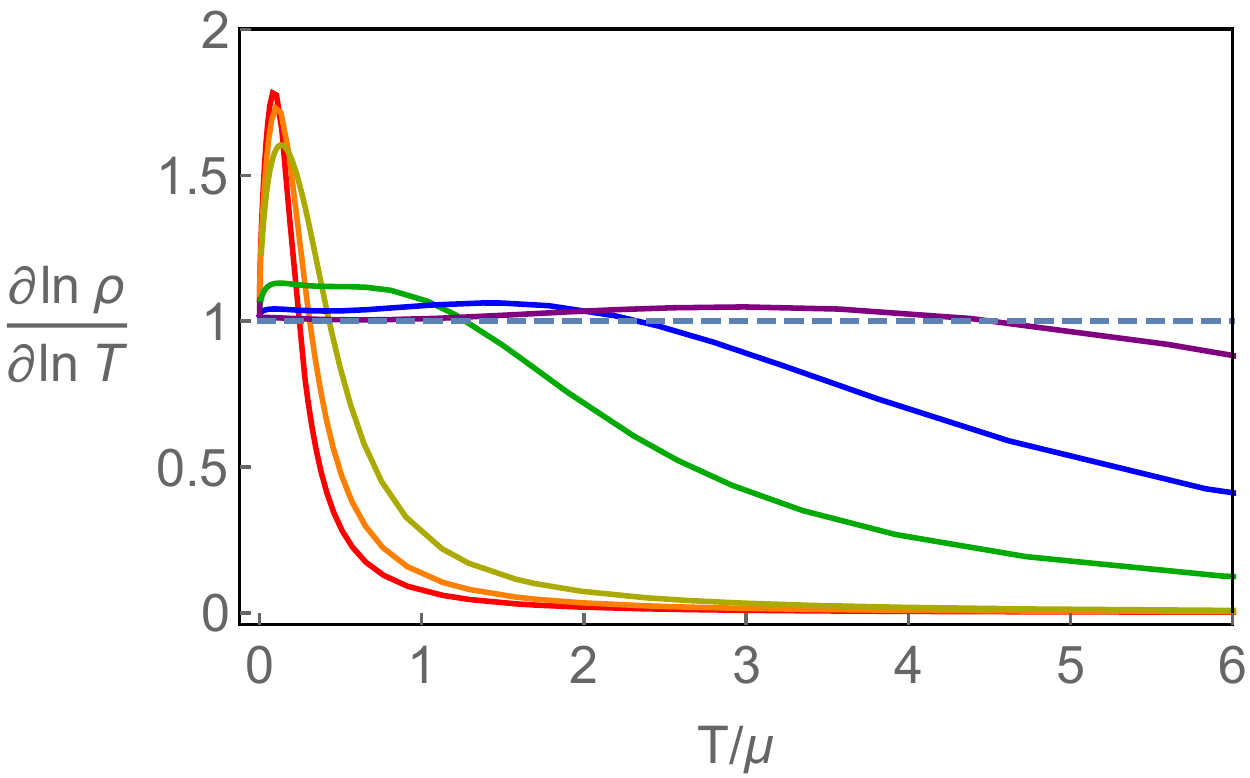} \label{fig:z3t1lrhoTpvm}}
 \caption{The temperature dependence of resistivity for $(\alpha, \beta, \gamma) = (-1/\sqrt{3}, -2/\sqrt{3}, \sqrt{3})$ or  $(z, \theta, \zeta) = (3, 1, -1)$. The color represents the momentum relaxation strength $k/\mu$: \{color($k/\mu$)\} = \{red(0.1), orange(0.5), yellow(1), green(5), blue(10), purple(20)\}. 
}\label{fig:z3t1vm}
\end{figure}

In this subsection, we will show how the momentum relaxation affects the linear-$T$ resistivity behavior in the finite temperature region. In Fig.~\ref{fig:z3t1vm}, we find that if the momentum relaxation becomes smaller, the temperature range of the linear-$T$ resistivity becomes shorter.  
The different colors in the figures represent different momentum relaxation strength: \{red, orange, yellow, green, blue, purple\} means $k/\mu= \{ 1/10,1/2, 1,5,10, 20$\}. 
From Fig.~\ref{fig:z3t1lrhoTvm}, one may think, for all $k/\mu$ we considered,  the resistivity looks linear in $T$ for certain range of $T$. However, in fact it is not. By carefully reading off the exponent $x$ in $\rho \sim T^x$ in Fig.~\ref{fig:z3t1lrhoTpvm}, we find that in most cases $x$ becomes quickly deviated from $1$ as $T$ increases from zero. Nevertheless, we stress all the curves in Fig.~\ref{fig:z3t1lrhoTpvm} go to $1$ as $T$ goes to zero. This confirms our numerics are consistent with the analytic formula \eqref{eq:res}. It turns out that the strong momentum relaxation is important to have robust linear-$T$ resistivity up to high temperature. For example, the range for the linear-$T$ resistivity is around up to $T/\mu \sim 5$ for $k/\mu  = 20$ and $T/\mu \sim 2$ for $k/\mu = 10$ as shown in Fig.~\ref{fig:z3t1lrhoTpvm}.


\section{Interpretations of numerical results}	\label{sec:5} 

\kyr{
In the previous section, we have shown the importance of large momentum relaxation to have a robust linear-$T$ resistivity at finite temperature. 
To have a better understanding on the mechanism for this observation, in this section, we want to answer the following two questions.
\begin{enumerate}
\item (section 5.1) To have a robust linear-$T$ resistivity, which term is important among two terms in \eqref{sigma}? We  will show it is the first term, so called the pair creation term. 
\item (section 5.2) What are the effects of $\alpha, \beta,$ and $\gamma$ in \eqref{ZandJV1} and \eqref{ZandJV2} on $x$ in $\rho \sim T^x$? We will show that
\begin{itemize}
\item{Increasing $\alpha$ or $\gamma$ $\Longrightarrow$ increasing $x$.}
\item{Increasing $\beta$ $\Longrightarrow$ dicreasing $x$.}
\end{itemize}
\end{enumerate}
}

\subsection{First term or second term?}

In general, the conductivity  \eqref{sigma} consists of two terms. The first term is called the pair creation term ($\sigma_{DC, pc}$) and the second term is called the dissipation term ($\sigma_{DC,diss}$) \cite{Gouteraux:2014hca}. For $d = 2$, \eqref{sigma} reads
\begin{equation}\label{d2sigma}
	\sigma_{DC} = \sigma_{DC, pc} + \sigma_{DC,diss} =   Z_H + \frac{q^2}{k^2  C_H J_H} \,.
\end{equation}
Thus, we may ask which term is responsible for the linear-$T$  resistivity at finite temperature.
%
\begin{figure}[]
\centering
     \subfigure[Class I, Eq.~\eqref{y1}]
     {\includegraphics[width=4.831cm]{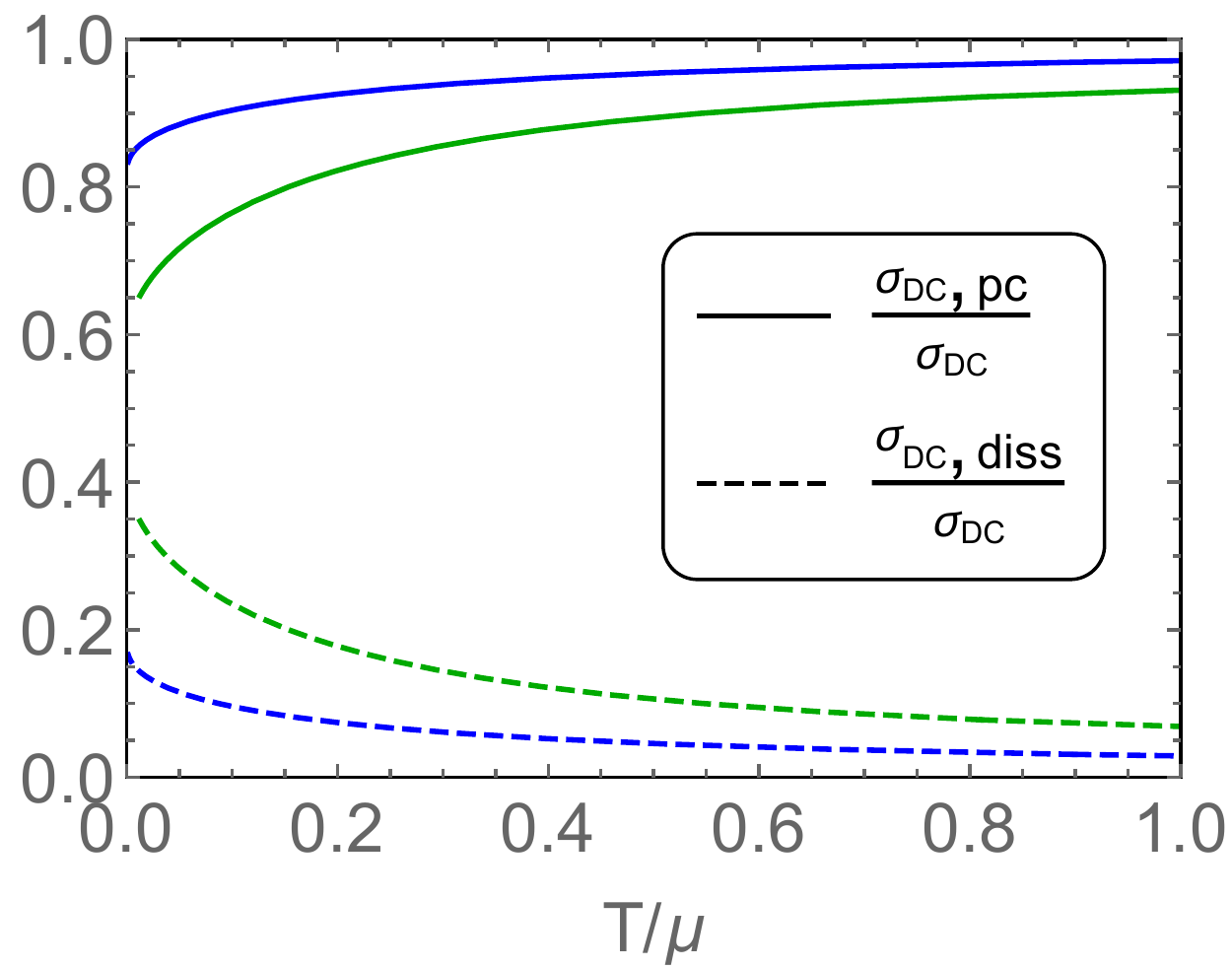} \label{fig:classIandcon}}
      \subfigure[Class II,  Eq.~\eqref{y2}]
     {\includegraphics[width=4.831cm]{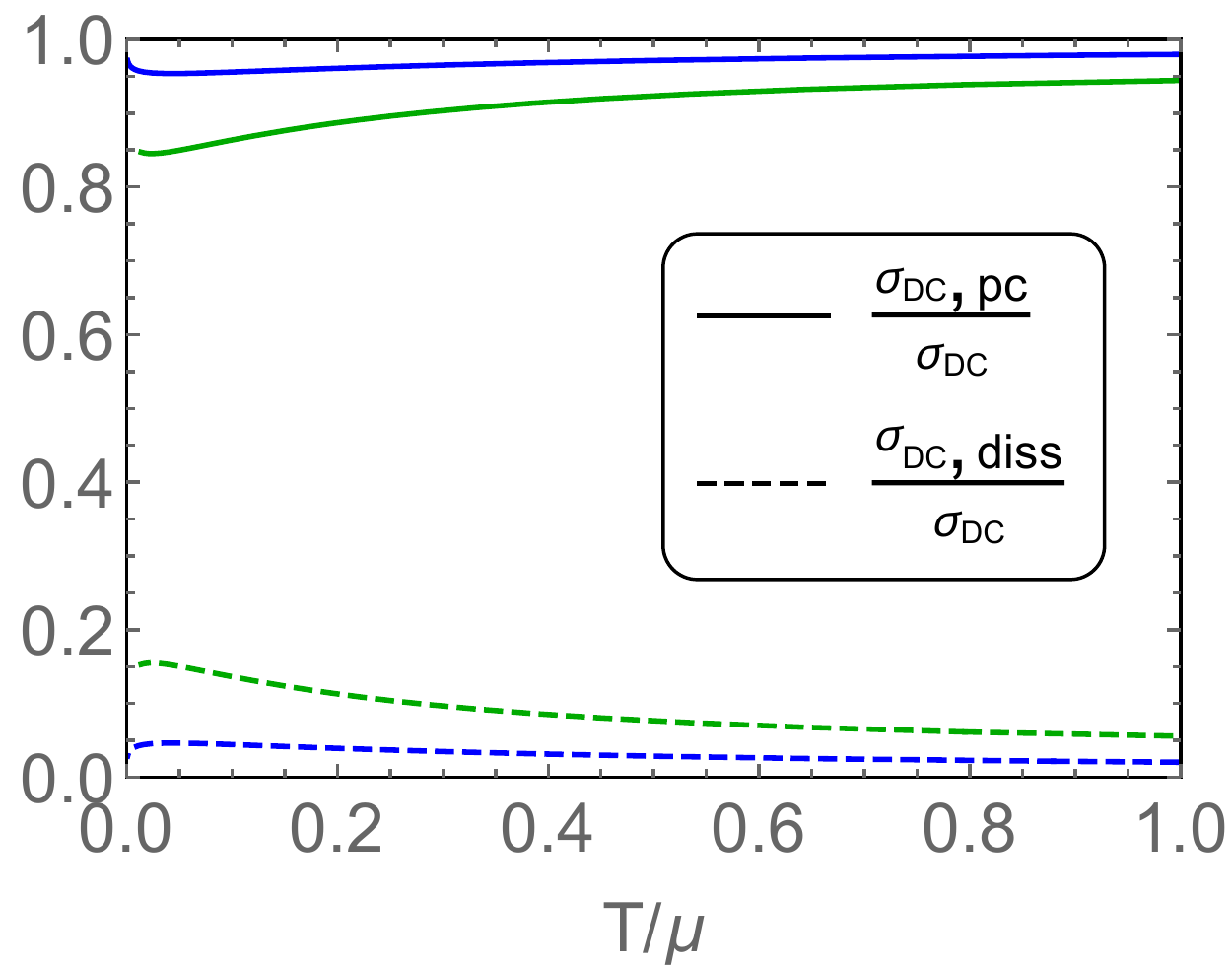} \label{fig:classIIandcon}}
      \subfigure[Class III,  Eq.~\eqref{y3}]
     {\includegraphics[width=4.831cm]{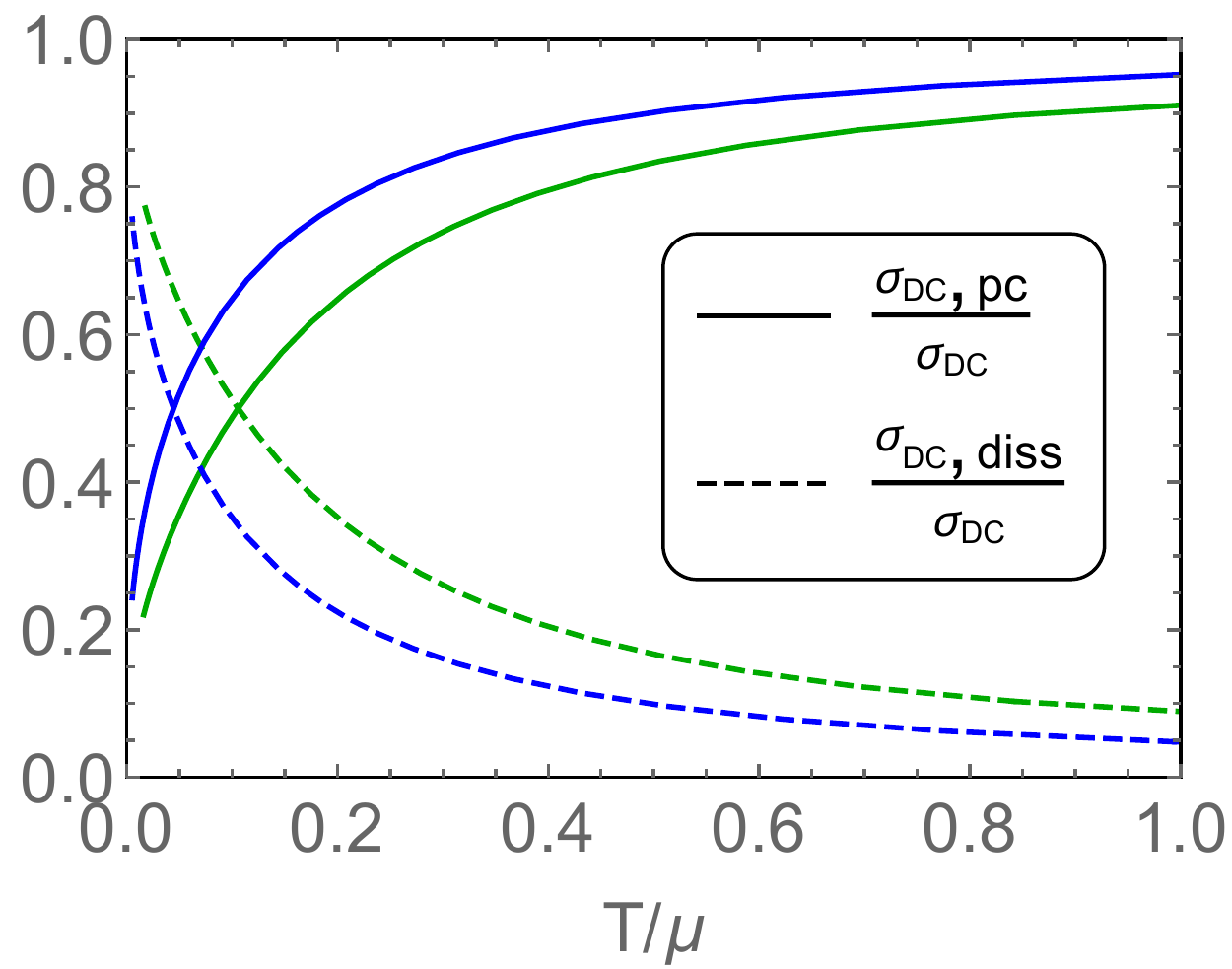} \label{fig:classIIIandcon}}
 \caption{The relative contributions of two terms to the conductivity in \eqref{d2sigma}.  {The green and blue curves represent $k/\mu = (5, 10)$ respectively.}}\label{fig:classandcon}
\end{figure} 
%

To answer this question we compare  $\frac{\sigma_{,DC,pc}}{\sigma_{DC}}$ and $\frac{\sigma_{DC,diss}}{\sigma_{DC}}$
 in Fig.~\ref{fig:classandcon}
for three representative cases\footnote{Here, we changed $z$ and $\zeta$ by one for comparison. In terms of $\alpha, \beta, \gamma$ our choice looks more complicated.} 
\begin{align}
&\mathrm{Class\ I}: \quad (\alpha, \beta, \gamma) = \left(-\frac{1}{\sqrt{3}}, -\frac{2}{\sqrt{3}}, \sqrt{3}\right)\,,\quad \left(z, \theta, \zeta\right) = (3, 1, -1)  \,, \label{y1} \\
&\mathrm{Class\ II}: \quad (\alpha,\beta,\gamma) = \left(\frac{1}{\sqrt{5}},-\frac{2}{\sqrt{5}},\frac{4}{\sqrt{5}}\right) \,, \quad (z,\theta,\zeta) = (4,1,-2)\,, \label{y2}  \\
&\mathrm{Class\ III}: \quad (\alpha,\beta,\gamma) = \left(\frac{1}{\sqrt{5}},-\frac{3}{\sqrt{5}},\frac{3}{\sqrt{5}}\right) \,, \quad (z,\theta,\zeta) = (4,1,-1)\,, \label{y3} 
\end{align}
where $(z, \theta, \zeta)$ can be calculated by using \eqref{classIexp}, \eqref{classIIexp}, \eqref{classIIIexp}.
We can determine which $(\alpha, \beta, \gamma)$ corresponds to which class by checking the equality and inequalities \eqref{con000},  \eqref{c2const}, and \eqref{constc3}.
For all cases, the resistivity is linear in $T$ in low temperature limit as shown in  \eqref{eq:res}. Class I \eqref{y1}, Class II \eqref{y2}, and Class III \eqref{y3} are shown in Fig.~\ref{fig:classIandcon}, Fiq.~\ref{fig:classIIandcon} and Fig.~\ref{fig:classIIIandcon} respectively. The green and blue curves represent the momentum relaxations, $k/\mu = 5$ and $k/\mu = 10$.

We find that, in general, at high temperature, the pair creation term contribute more dominantly. It can be understood by more active pair creation at high temperature. 
At strong momentum relaxation, naively, one may think that the second term is always suppressed because there is $k^2$ factor in the denominator. However, this is not obvious because all the other factors $q, C_H$, and $J_H$ are also implicitly functions of $k$. Indeed, we find that the dissipation term is dominant at low temperature in Class III in Fig.~\ref{fig:classIIIandcon}.\

In particular, Fig.~\ref{fig:classIandcon} is for Eq.~\eqref{y1}, which exhibits a robust linear-$T$ resistivity. In this case, the larger momentum relaxation is, the more pair creation term is dominant. Thus, we find that the pair creation term, the horizon value of $Z$ ($Z_H$), is responsible for the linear-$T$ resistivity at finite temperature. 

Furthermore, from  Fig. \ref{fig:classIIIandcon}, we may understand why class III case is hard to exhibit the robust linear-$T$ resistivity at finite temperature. As temperature increases, the dominant mechanism for conductivity is changed: at low temperature, the dissipation term dominates while at high temperature the pair creation term dominates. It will be more difficult to have a universal physics from two different mechanism.\footnote{If the momentum relaxation is weak, the dissipation term is always dominant so there is no crossing in Fig. \ref{fig:classIIIandcon}. We thank Blaise Gout\'{e}raux for pointing this out.}


\subsection{$\alpha, \beta, \gamma$ dependence} \label{sec43}

In the previous subsection, we have investigated the effect of the momentum relaxation on the resistivity in finite temperature regime. In this subsection we want to investigate the effect of $\{ \alpha, \beta, \gamma \}$ or $ \{ z,\theta,\zeta \}$ on the resistivity. 

Because we found that, in general, the large momentum relaxation is important to have a robust linear-$T$ resistivity, here we fix the momentum relaxation to be large, say $k/\mu =10$. For a systematic study we first need to choose a potential in \eqref{ZandJV2}. Thus, we have three cases depending on the sign of $\theta$ 
\begin{equation}
\theta >0 \,, \quad  \theta = 0 \,,  \quad \theta <0 \,.
\end{equation}
This classification is equivalent to
\begin{equation}
\alpha <0 \,, \quad  \alpha = 0 \,,  \quad \alpha >0 \,,
\end{equation}
respectively because $\alpha \kappa = - \frac{2\theta}{d}$ with a positive dilaton  $\phi$ ($\kappa >0$). 

\begin{figure}[]
\centering
      \subfigure[$\alpha = -\frac{14}{10}\frac{1}{\sqrt{3}}$]
     {\includegraphics[width=4.831cm]{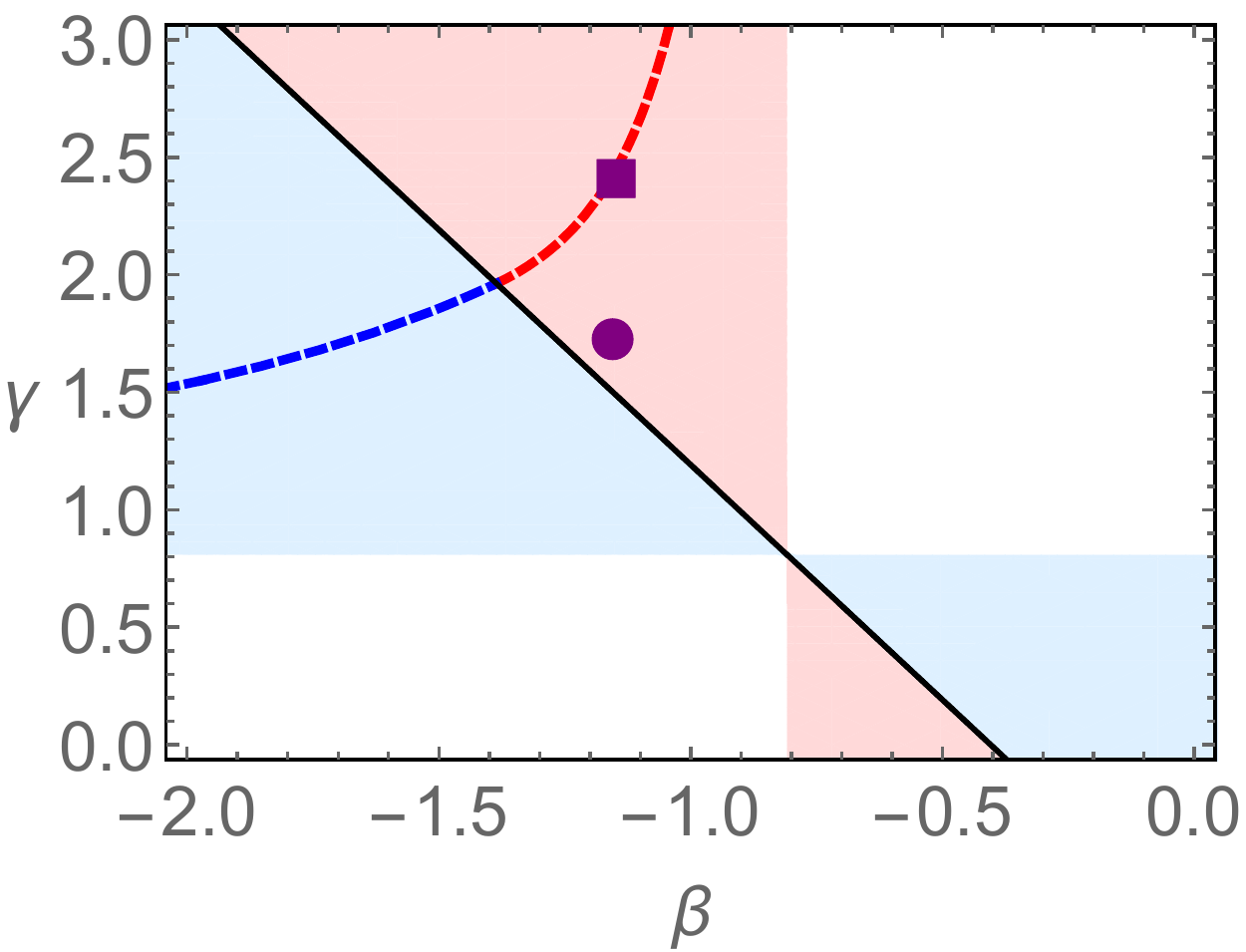} \label{fig:pthregiona}}
      \subfigure[$\alpha = -\frac{1}{\sqrt{3}}$]
     {\includegraphics[width=4.831cm]{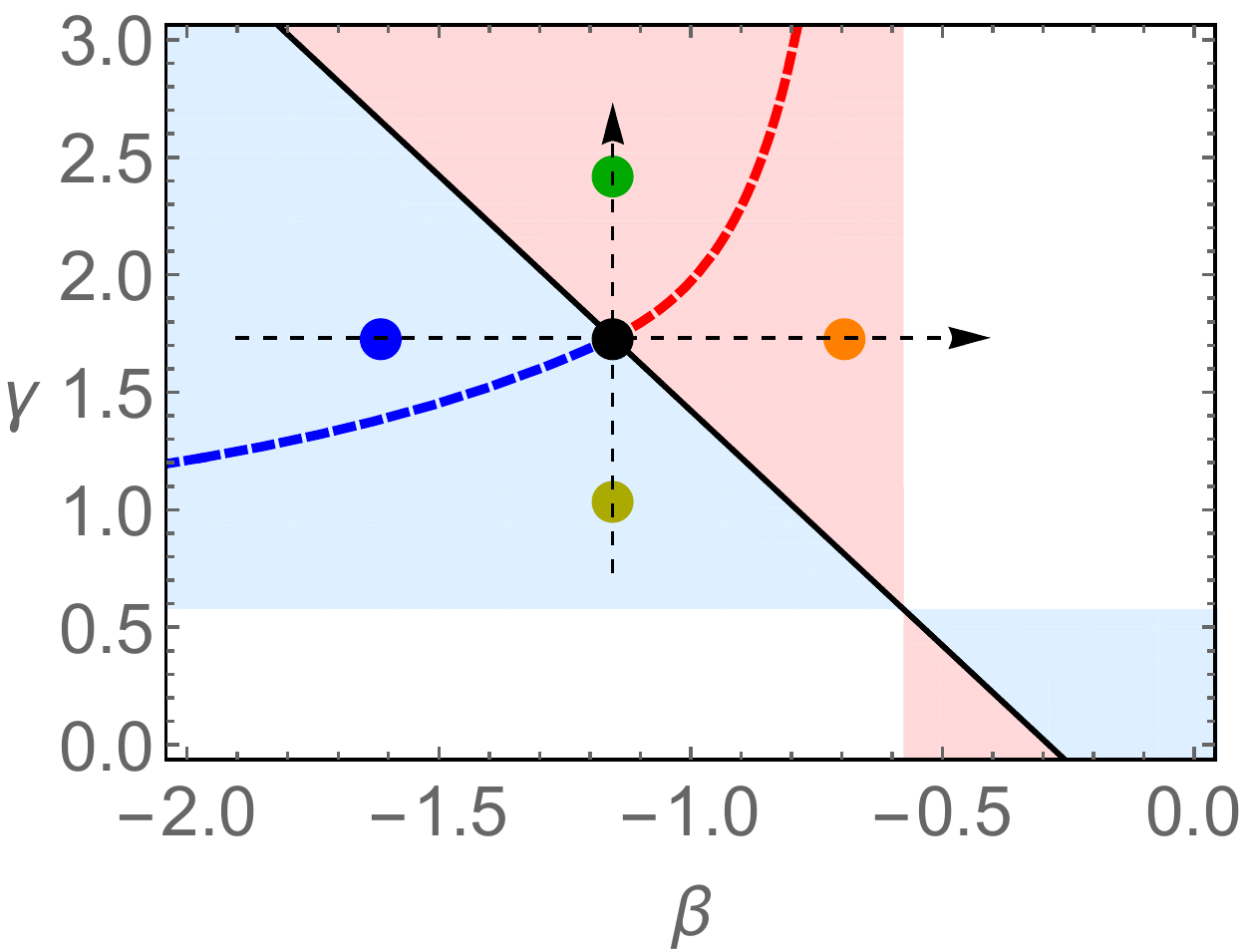} \label{fig:pthregionb}}
     \subfigure[$\alpha = -\frac{6}{10}\frac{1}{\sqrt{3}}$]
     {\includegraphics[width=4.831cm]{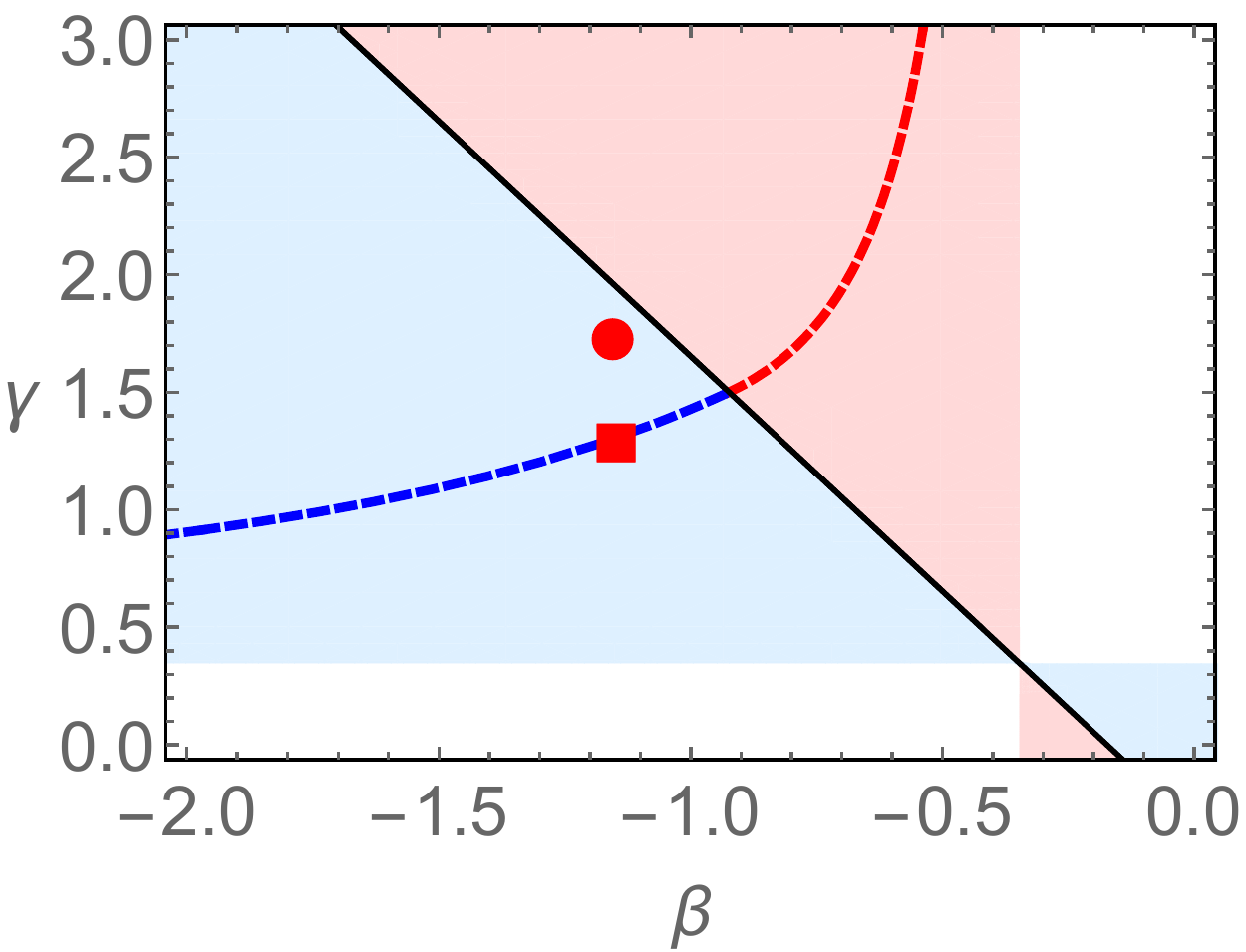} \label{fig:pthregionc}}
 \caption{Allowed region of $(\beta, \gamma)$ for a given $\alpha <0$. 
 In Fig. \ref{abgdepp3} we show the resistivity for the parameters corresponding to the dots in (a), (b), and (c). See the items 1-5 in sec,~\ref{sec43} for the meanings of colors, lines, and dots. In Fig. \ref{shift} we  show the resistivity for the squares in (a) and (c).
}\label{fig:pthregion}
\end{figure}

\begin{figure}[]
\centering
      \subfigure[$\alpha$ dependence]
     {\includegraphics[width=4.831cm]{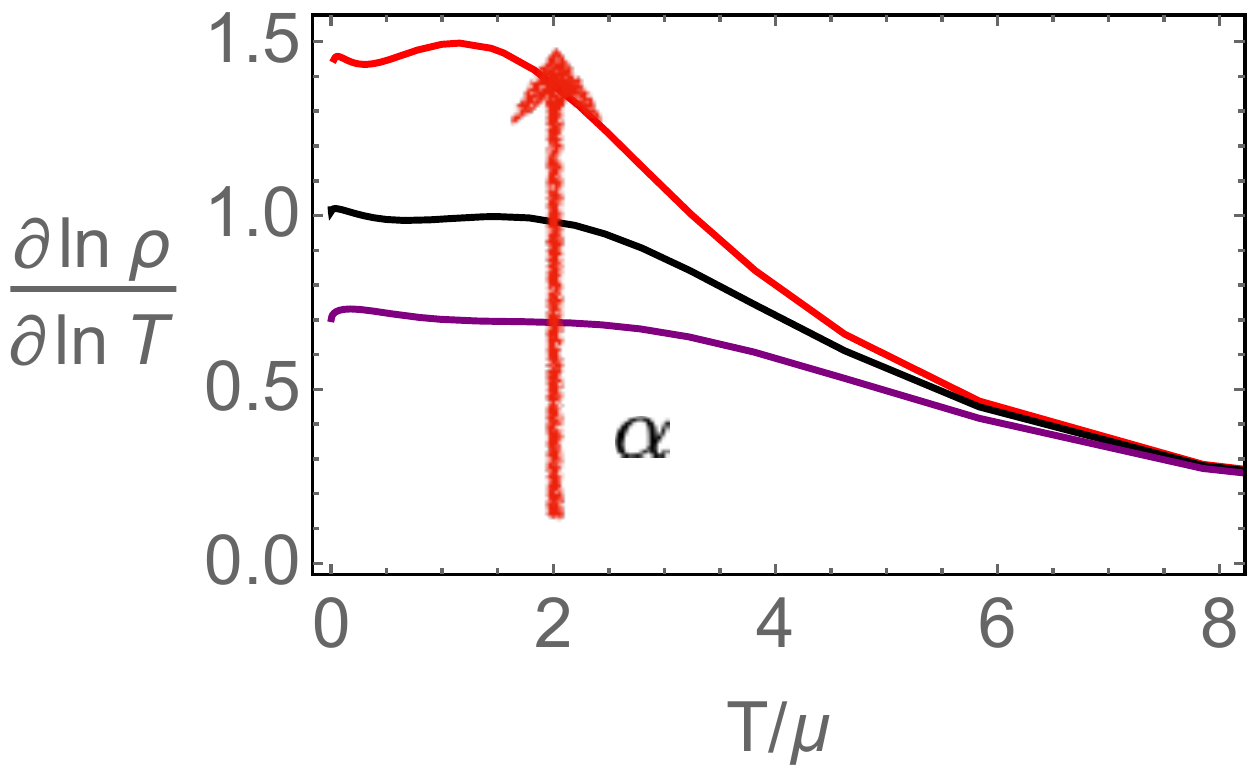} \label{abgdepp3a}}
      \subfigure[$\beta$ dependence]
     {\includegraphics[width=4.831cm]{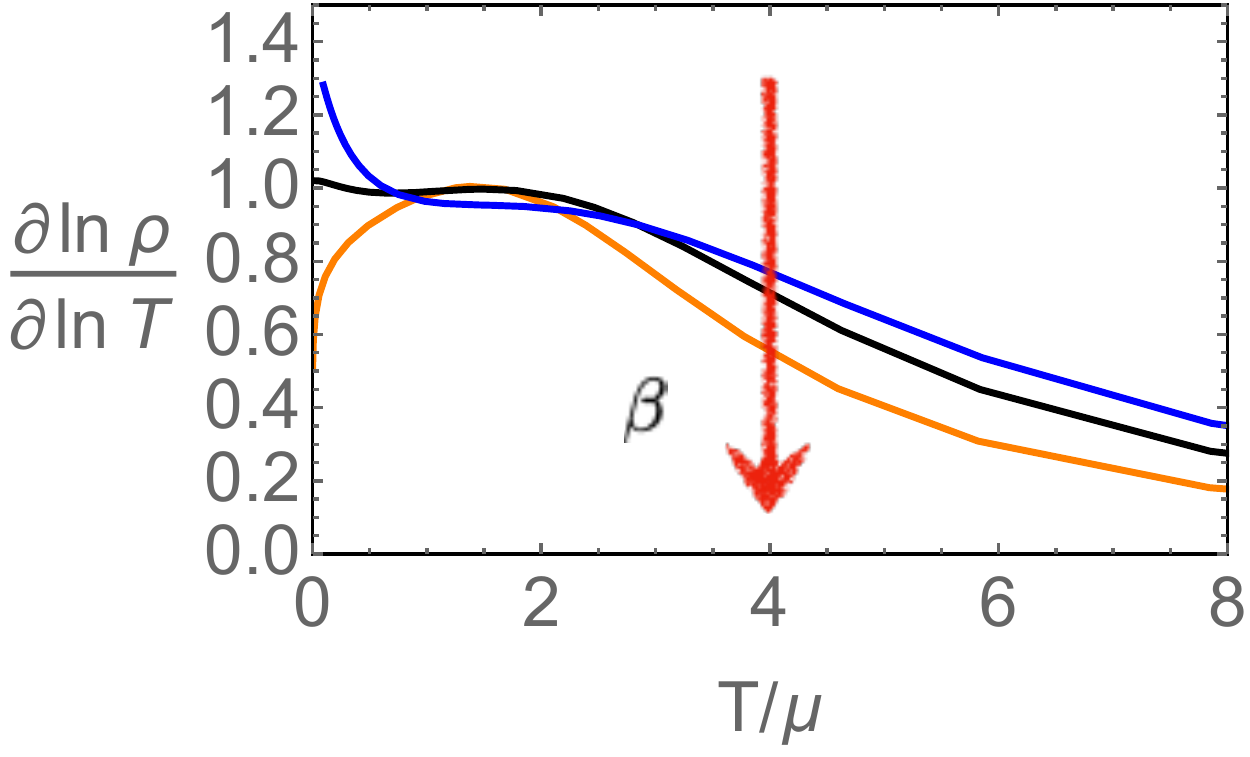} \label{abgdepp3b}}
     \subfigure[$\gamma$ dependence]
     {\includegraphics[width=4.831cm]{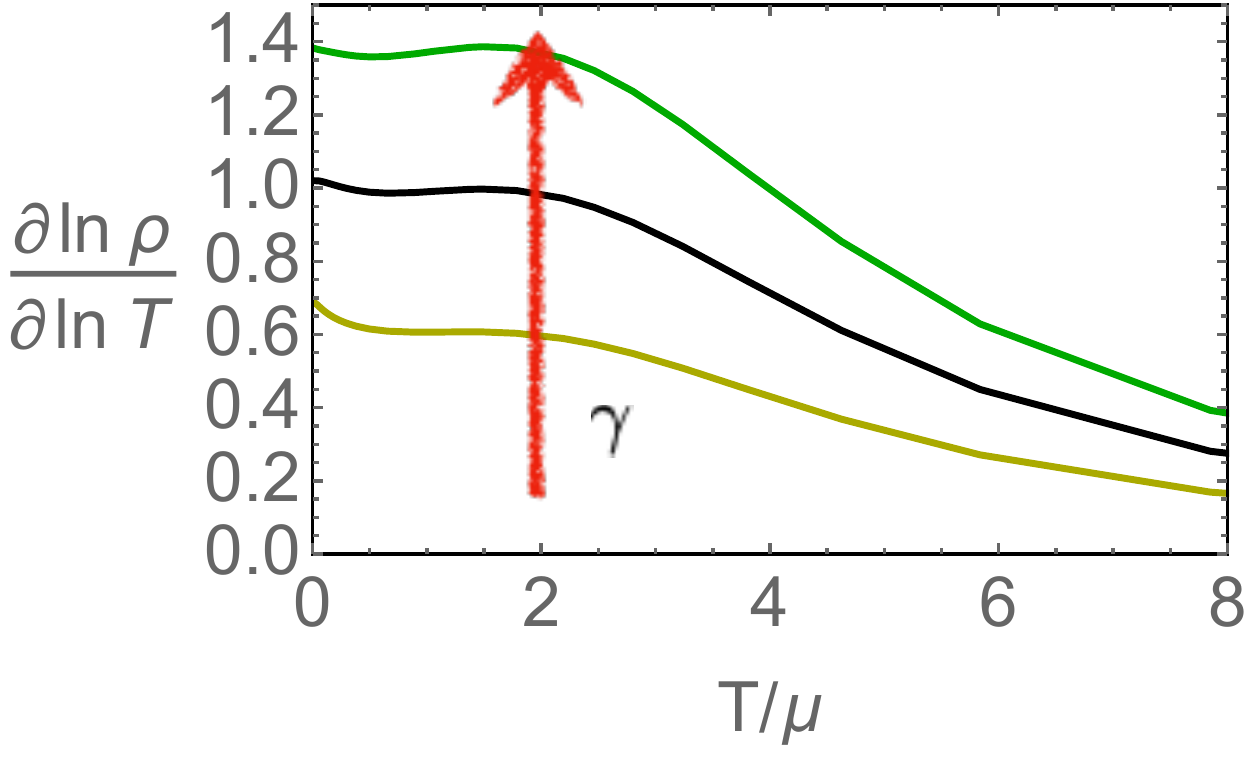} \label{abgdepp3c}}
 \caption{
 The exponent $x$ in $\rho \sim T^x$. The colors of the curves are chosen to be the same as the colors of the dots in Fig.~\ref{fig:pthregion}.
 }\label{abgdepp3}
\end{figure}

\begin{figure}[]
\centering
     {\includegraphics[width=7cm]{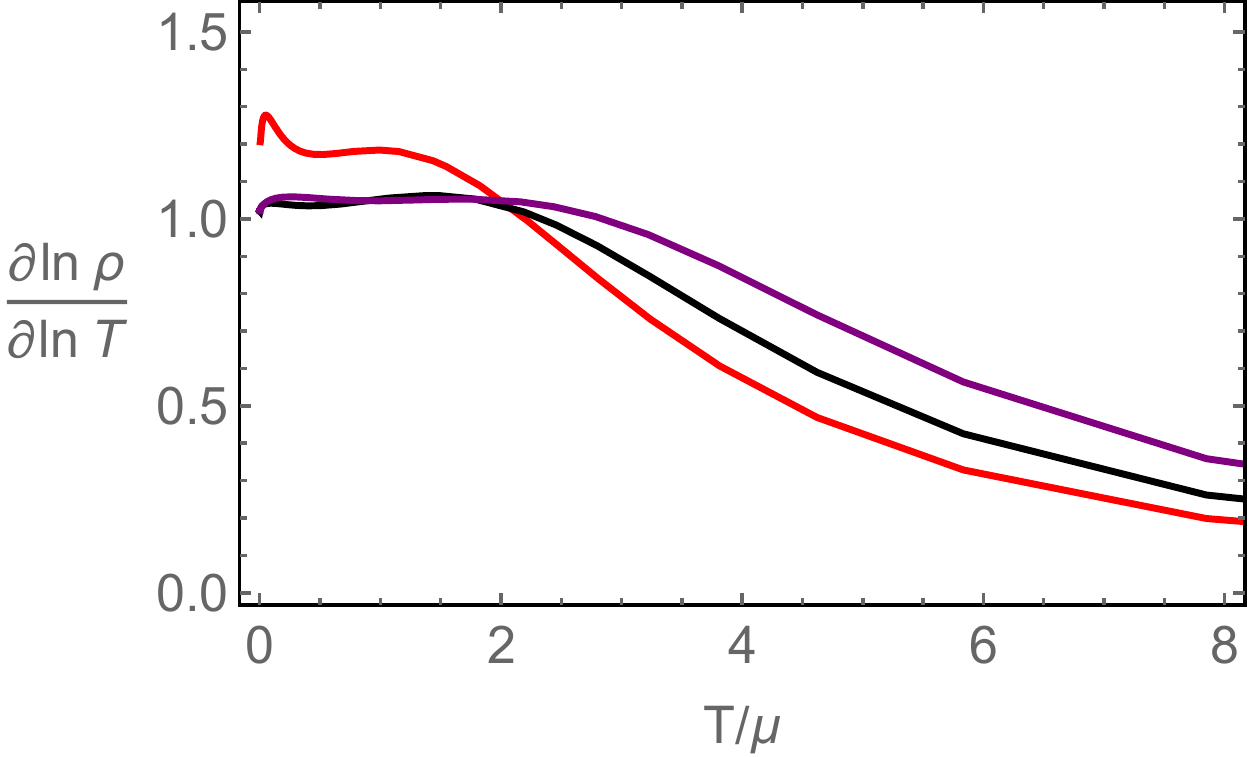} \label{}}
 \caption{The exponent $x$ in $\rho \sim T^x$. The black curve corresponds to the black dot in Fig.~\ref{abgdepp3a}. The purple (red) curve corresponds to the square in Fig.~\ref{abgdepp3a} (Fig.~\ref{abgdepp3c}).}\label{shift}
\end{figure}

\begin{figure}[]
\centering
     \subfigure[$\alpha = \frac{6}{10}\frac{1}{3\sqrt{3}}$]
     {\includegraphics[width=4.831cm]{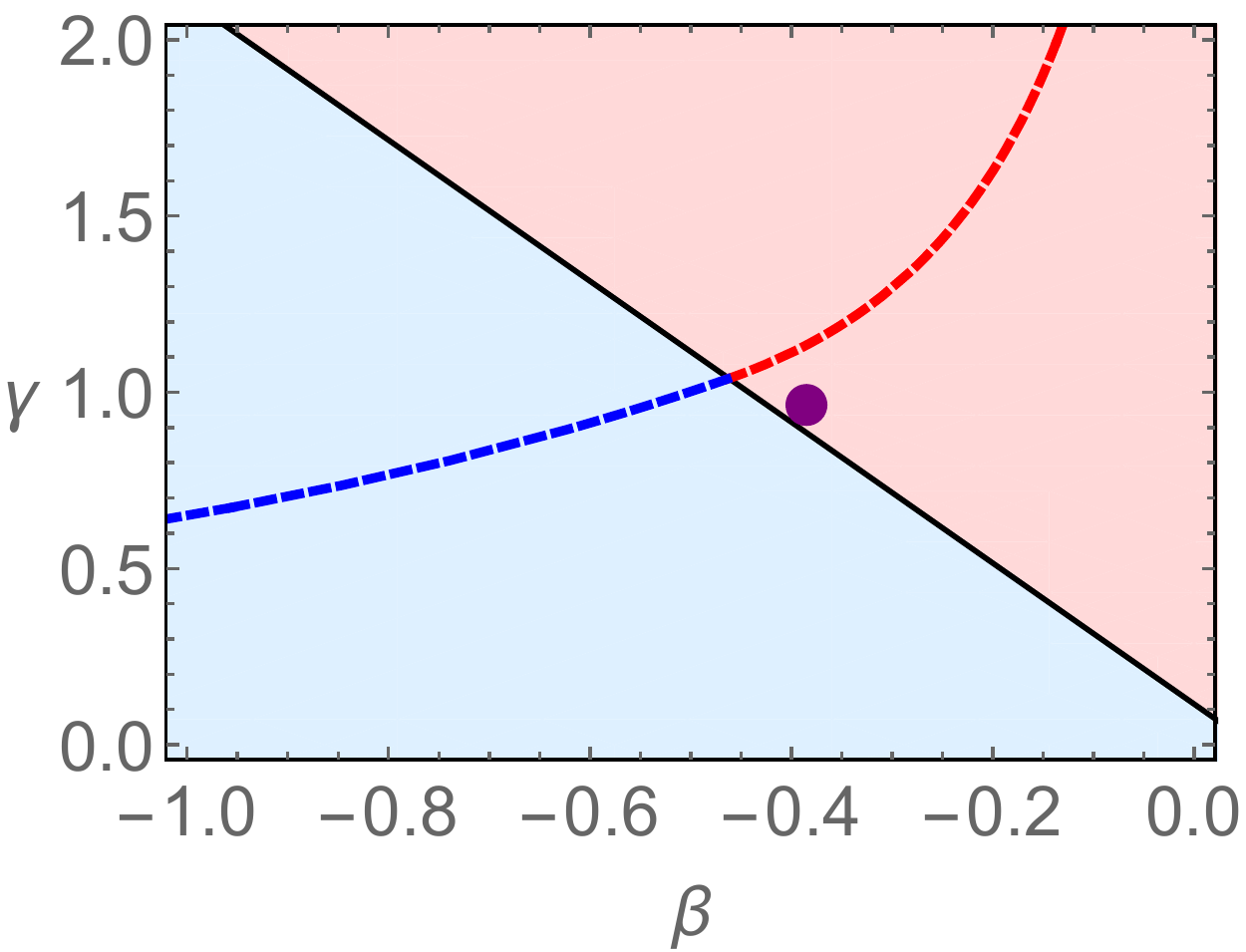} \label{fig:nthregion1}}
      \subfigure[$\alpha = \frac{1}{3\sqrt{3}}$]
     {\includegraphics[width=4.831cm]{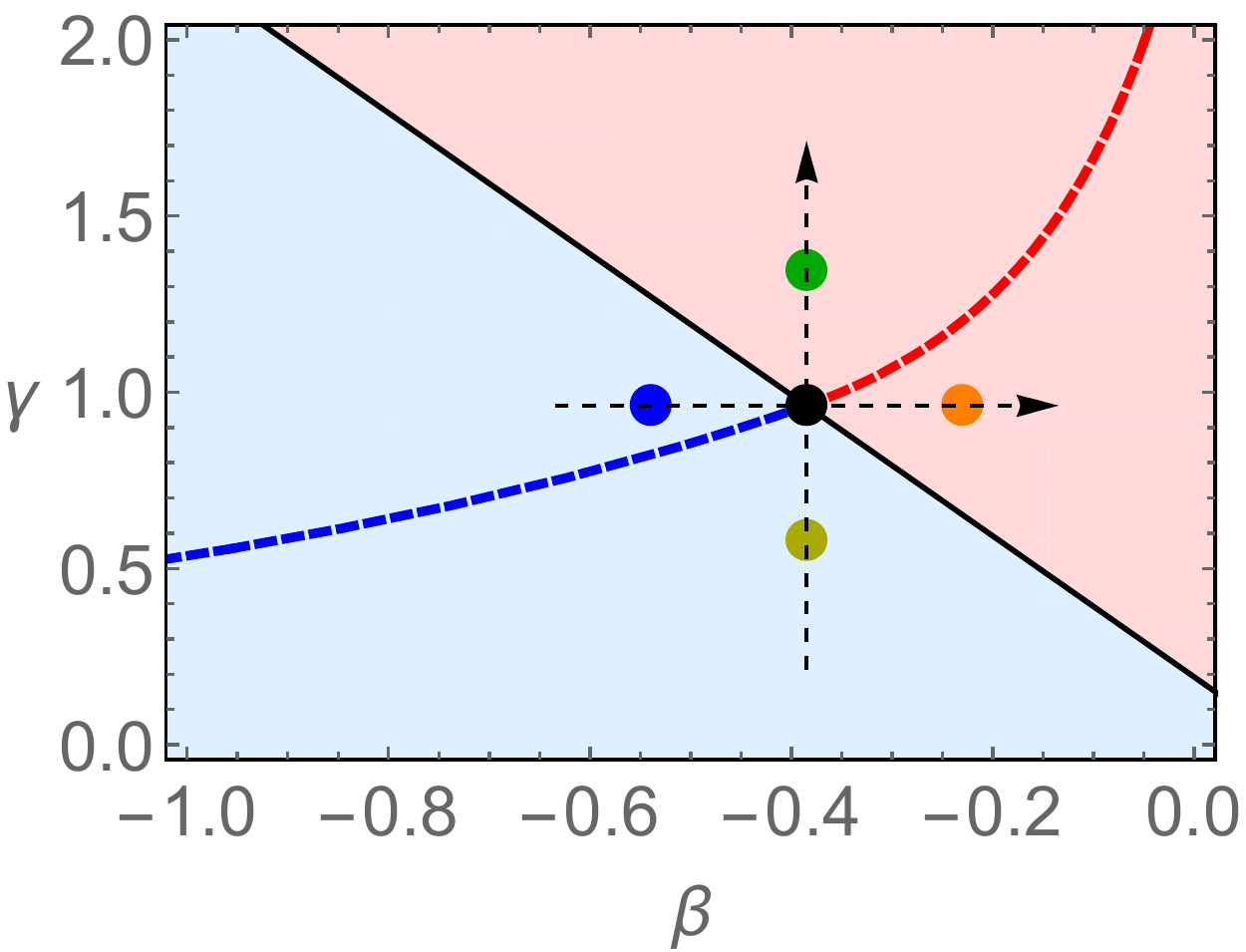} \label{fig:nthregion2}}
      \subfigure[$\alpha = \frac{14}{10}\frac{1}{3\sqrt{3}}$]
     {\includegraphics[width=4.831cm]{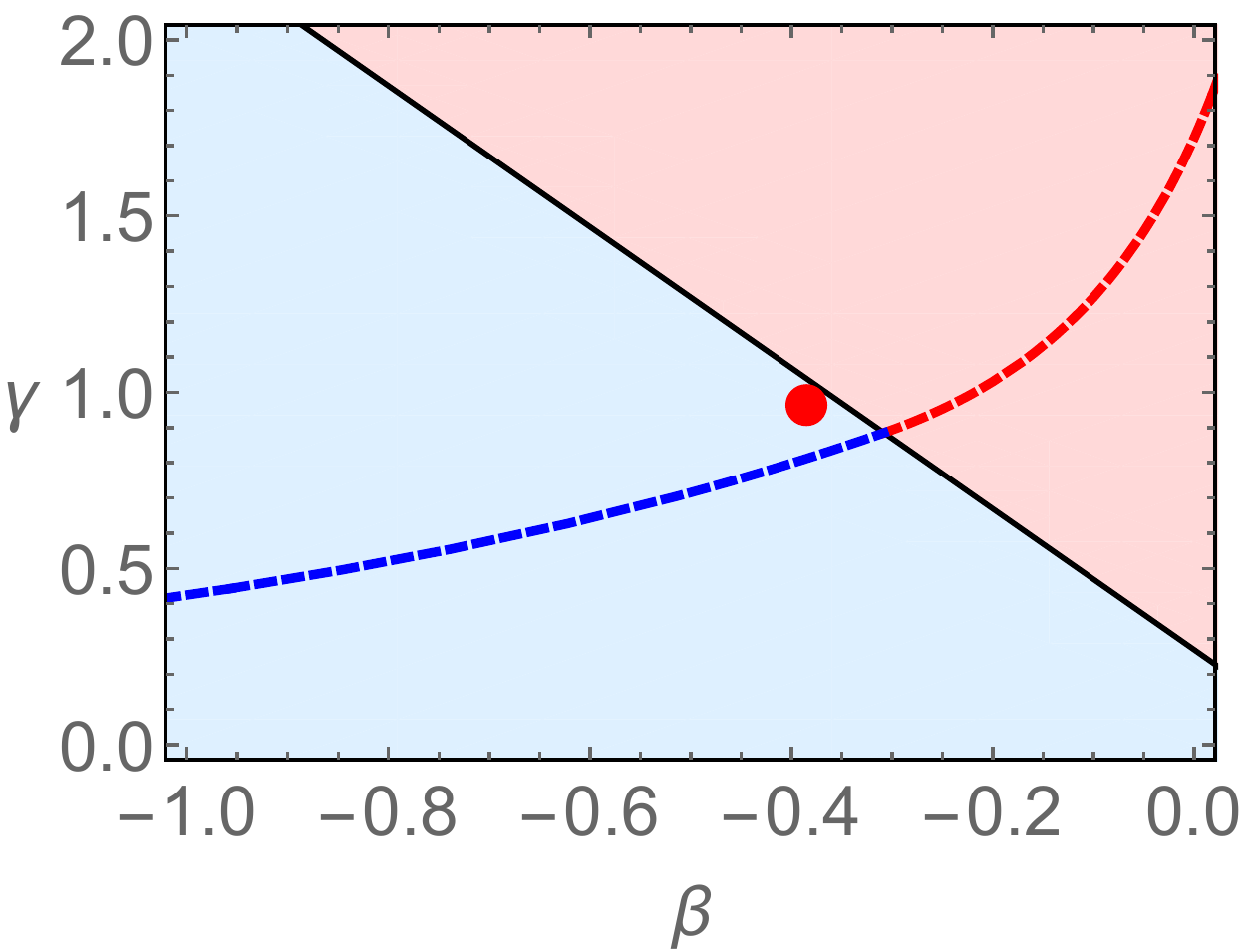} \label{fig:nthregion3}}
 \caption{Allowed $(\beta, \gamma)$ region for a given $\alpha > 0$. In Fig. \ref{abgdepp1} we show the resistivity for the parameters corresponding to the dots in (a), (b), and (c).
 See the items 1-5 in sec,~\ref{sec43} for the meanings of colors, lines, and dots. 
 }\label{fig:nthregion}
\end{figure}

\begin{figure}[]
\centering
      \subfigure[$\alpha$ dependence]
     {\includegraphics[width=4.831cm]{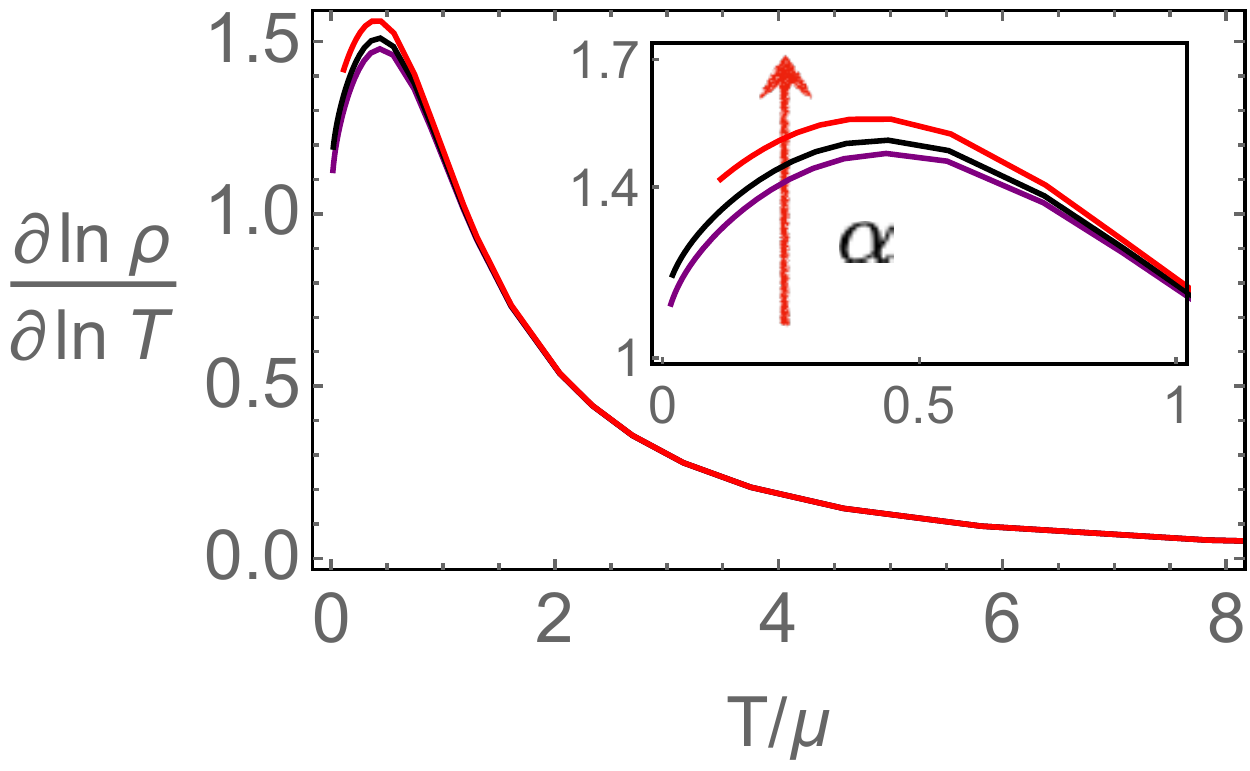} \label{abgdepp1a}}
      \subfigure[$\beta$ dependence]
     {\includegraphics[width=4.831cm]{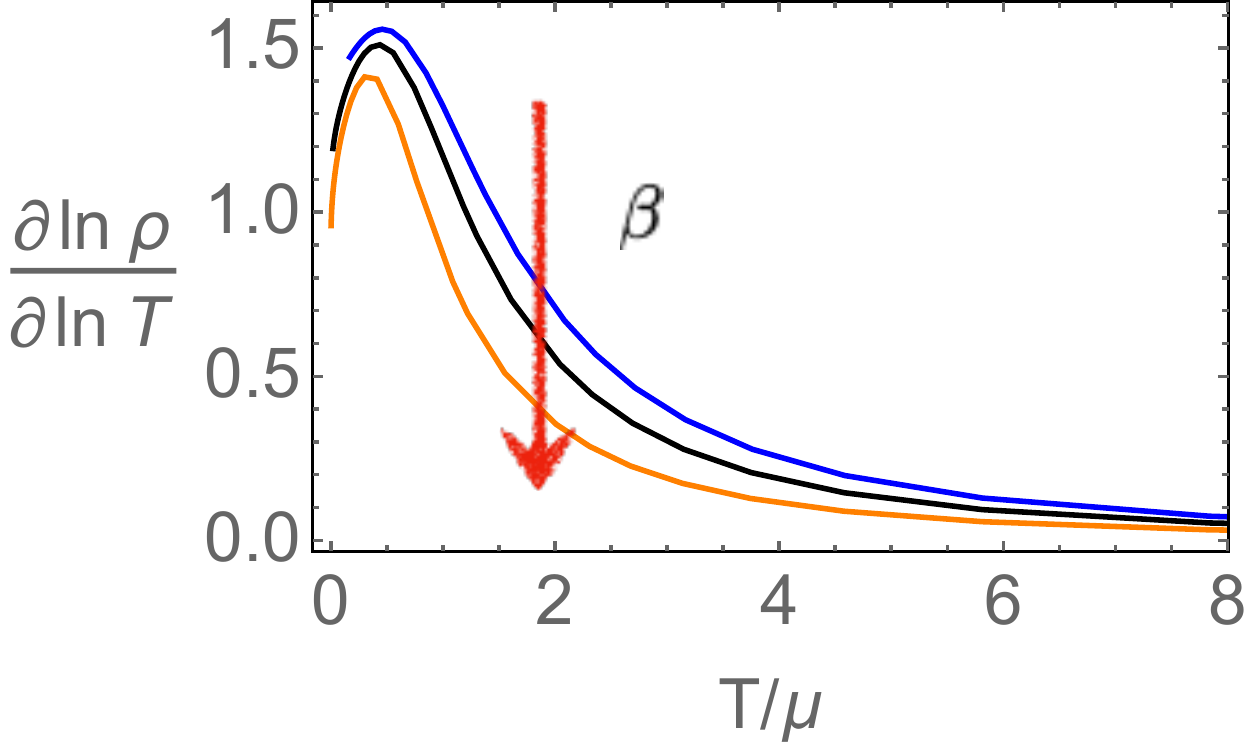} \label{abgdepp1b}}
      \subfigure[$\gamma$ dependence]
     {\includegraphics[width=4.831cm]{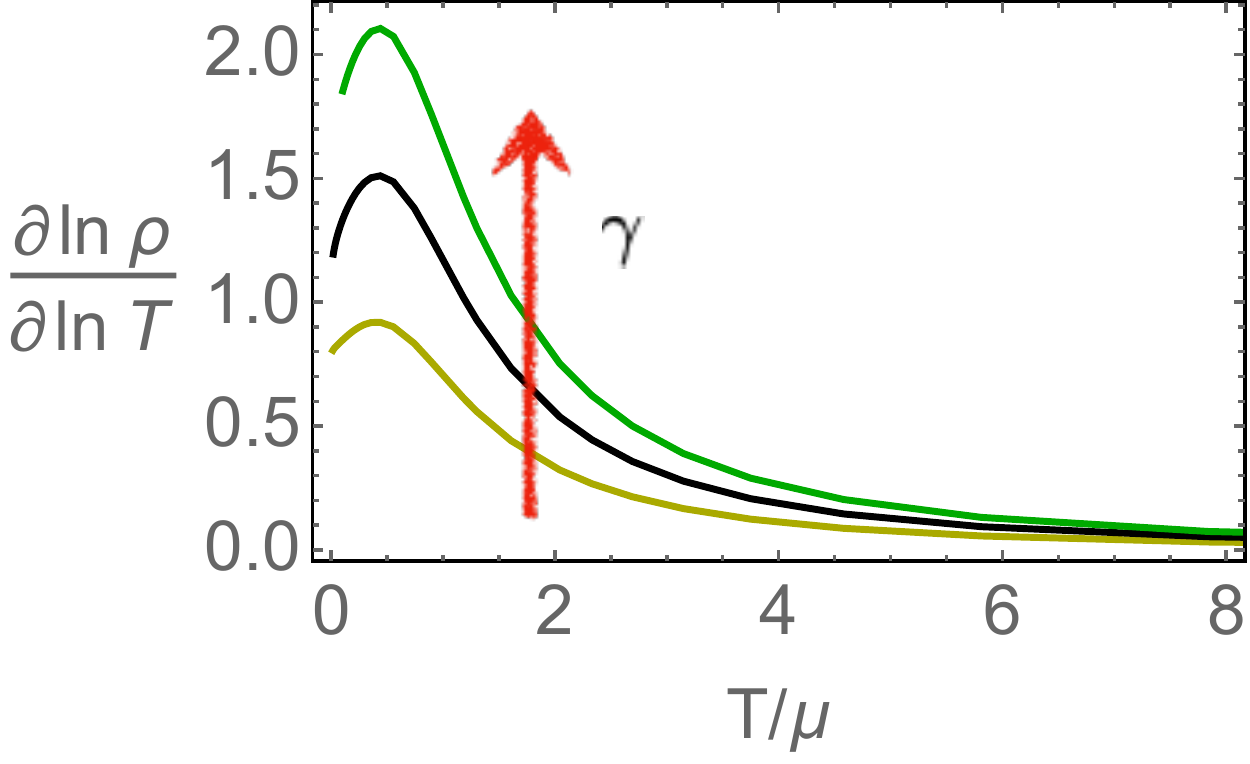} \label{abgdepp1c}}
 \caption{The exponent $x$ in $\rho \sim T^x$. The colors of the curves are chosen to be the same as the colors of the dots in Fig.~\ref{fig:nthregion}. }\label{abgdepp1}
\end{figure}

\begin{figure}[]
\centering
     {\includegraphics[width=5cm]{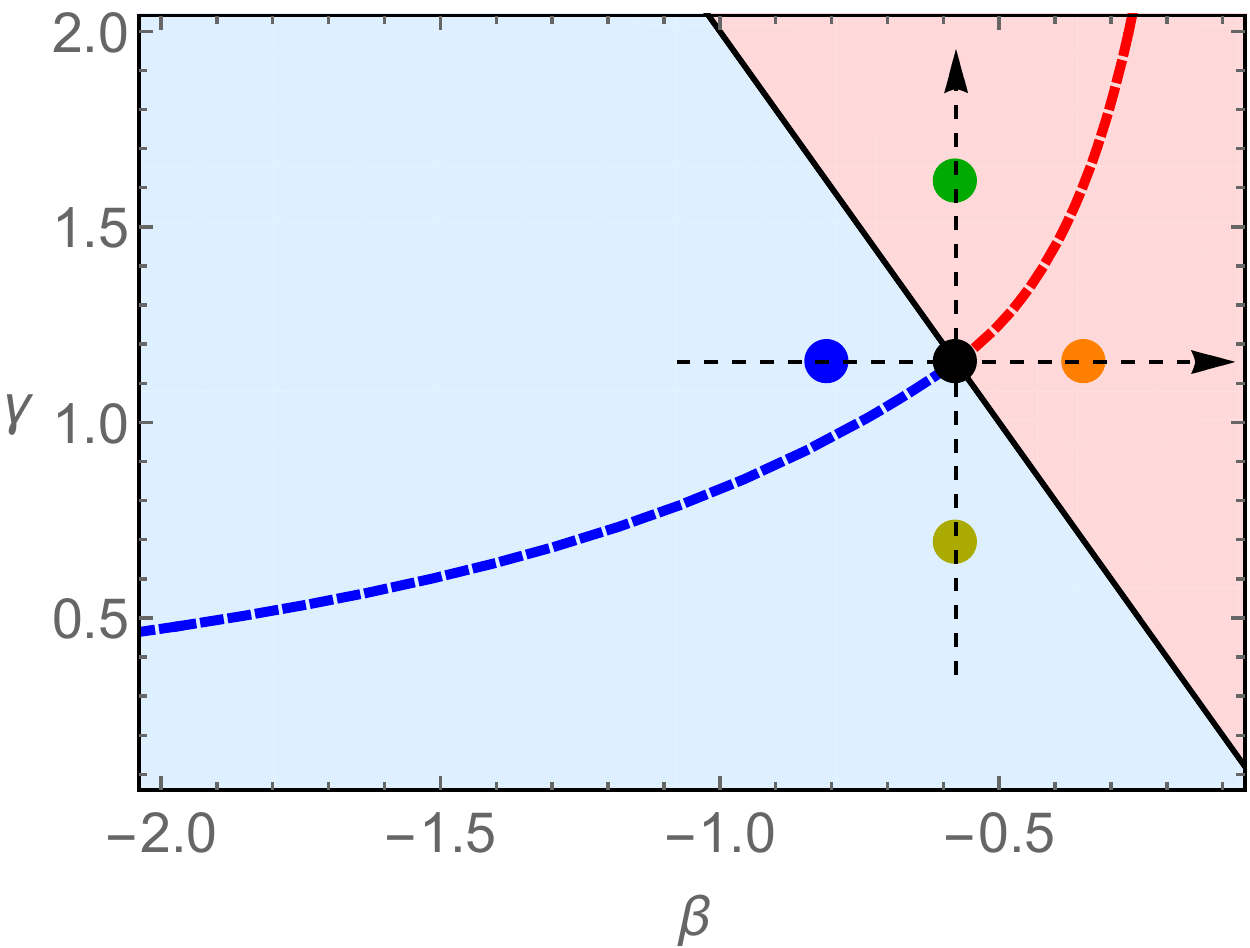} \label{}}
 \caption{Allowed region of $(\beta, \gamma)$ for $\alpha = 0$. In Fig. \ref{bgdepp2} we show the resistivity for the parameters corresponding to the dots.  
 See the items 1-5 in sec,~\ref{sec43} for the meanings of colors, lines, and dots.
 }\label{fig:zthregion}
\end{figure}

\begin{figure}[]
\centering
      \subfigure[$\beta$ dependence]
     {\includegraphics[width=5cm]{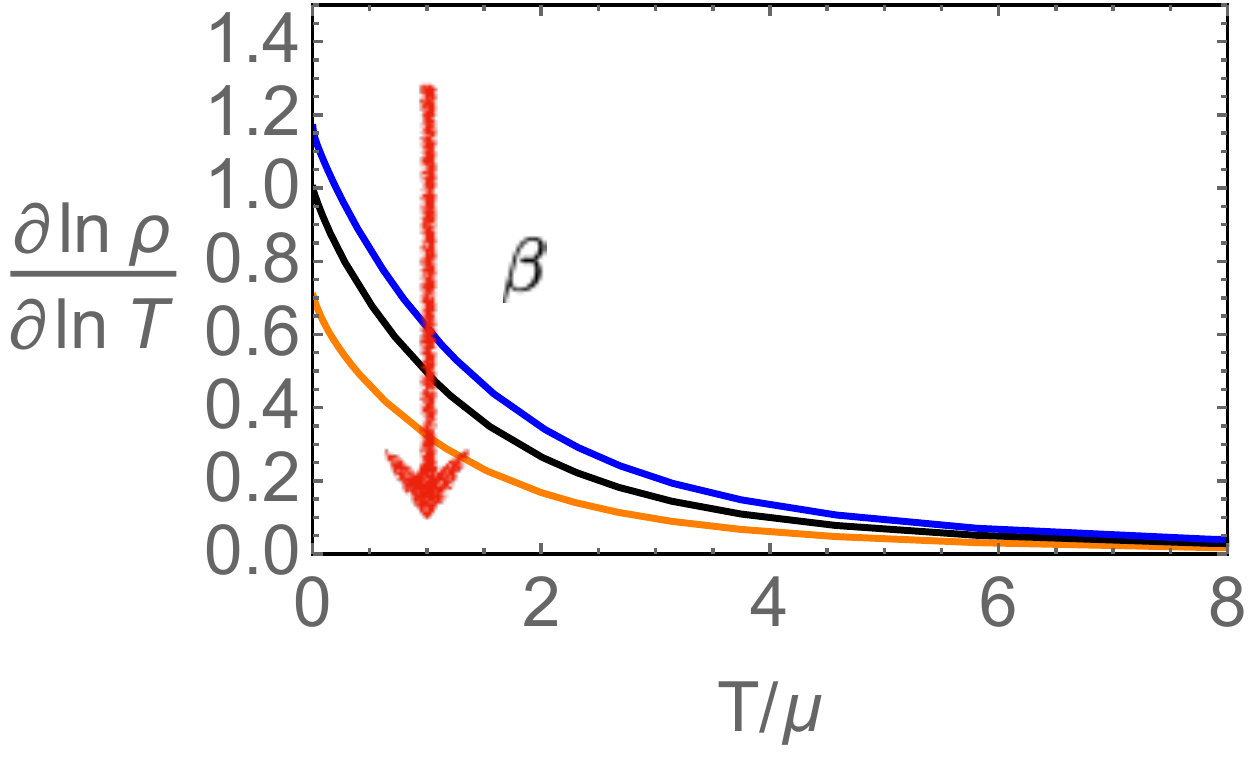} \label{bgdepp2a}}
      \subfigure[$\gamma$ dependence]
     {\includegraphics[width=5cm]{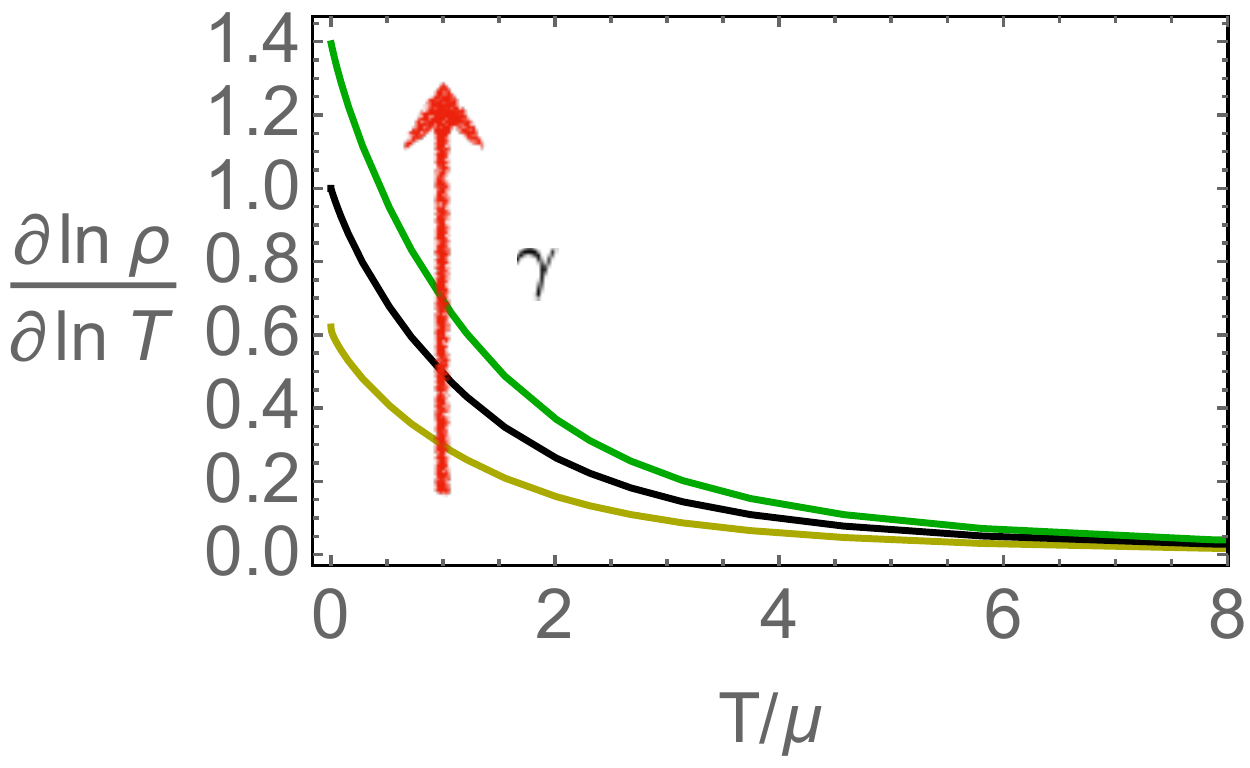} \label{bgdepp2b}}
 \caption{The exponent $x$ in $\rho \sim T^x$. The colors of the curves are chosen to be the same as the colors of the dots in Fig.~\ref{fig:zthregion}.
}\label{bgdepp2}
\end{figure}

For a given potential we have three classes Class I,II, and III explained in sec. ~\ref{sec22}. The classes are determined by the parameter range. This parameter range and the effect of $\alpha,\beta,\gamma$ on resistivity are best explained by figures, from Fig.~\ref{fig:pthregion} to Fig.~\ref{bgdepp2}.

Because there are common features in a class of figures (Fig.~\ref{fig:pthregion}, Fig.~\ref{fig:nthregion}, Fig.~\ref{fig:zthregion}) and another class of figures (Fig.~\ref{abgdepp3}, Fig.~\ref{abgdepp1}, Fig.~\ref{bgdepp2})  we explain them here for all of them.
\begin{enumerate}
\item (Fig.~\ref{fig:pthregion}, Fig.~\ref{fig:nthregion},  Fig.~\ref{fig:zthregion}) show allowed region of $(\beta, \gamma)$ for a given $\alpha$. The red region, blue region, and the black line between them correspond to class II, class III, and class I  respectively.  
\item The dotted line in blue and red corresponds to the parameters giving the linear-$T$ resistivity in low temperature limit. 
Let us imagine that we file up the dashed lines for every $\alpha$. Then, the blue(red) dashed lines form a blue(red) surface in $(\alpha, \beta, \gamma)$ space.  This surface is nothing but the surface in Fig.~\ref{fig:wregion}. In other words, the dashed lines in figures (Fig.~\ref{fig:pthregion}, Fig.~\ref{fig:nthregion}, Fig.~\ref{fig:zthregion}) are cross-section of Fig.~\ref{fig:wregion} at a given $\alpha$.   The region above (below) the dashed line corresponds to $x>1$ ($x<1$), where $x$ is defined in the relation $\rho \sim T^x$  in low temperature limit.  
\item Therefore, the  black dot in (Fig.~\ref{fig:pthregion}, Fig.~\ref{fig:nthregion}, Fig.~\ref{fig:zthregion}) gives the linear-$T$ resistivity as a class I case. Because all the other color dots do not belong to the dotted line they do not give the linear-$T$ resistivity in low temperature limit. We chose these deviated points to investigate the effect of $\alpha, \beta, \gamma$.
\item $\alpha$ increase as the purple dot $\rightarrow$ the black dot $\rightarrow$ the red dot in (Fig.~\ref{fig:pthregion} and Fig.~\ref{fig:nthregion}). This was shown as an arrow in (Fig.~\ref{abgdepp3a}, Fig.~\ref{abgdepp1a}).   
\item $\beta$ increase as the blue dot $\rightarrow$ the black dot $\rightarrow$ the orange dot in (Fig.~\ref{fig:pthregion}, Fig.~\ref{fig:nthregion}, Fig.~\ref{fig:zthregion}). This was shown as an arrow in (Fig.~\ref{abgdepp3b}, Fig.~\ref{abgdepp1b}, Fig.~\ref{bgdepp2a}).
\item  $\gamma$ increase as the dark yellow dot $\rightarrow$ the black dot $\rightarrow$ the green dot in (Fig.~\ref{fig:pthregion}, Fig.~\ref{fig:nthregion}, Fig.~\ref{fig:zthregion}). This was shown as an arrow in (Fig.~\ref{abgdepp3c}, Fig.~\ref{abgdepp1c}, Fig.~\ref{bgdepp2b}). 
\item (Fig.~\ref{abgdepp3}, Fig.~\ref{abgdepp1}, Fig.~\ref{bgdepp2}) shows the exponent $x$ in $\rho \sim T^x$. The colors of the curves are chosen to be the same as the colors of the dots in (Fig.~\ref{fig:pthregion}, Fig.~\ref{fig:nthregion}, Fig.~\ref{fig:zthregion}).
Recall that the region above (below) the dashed line corresponds to $x>1$ ($x<1$), where $x$ is defined in the relation $\rho \sim T^x$  in low temperature limit.  Our results show that this tendency remain at finite temperature in general.
\item 		As a consistency check of our numerics, we  have confirmed that the numerical values of $x$ in the limit  $T/\mu \rightarrow 0 $ for every curve in (Fig.~\ref{abgdepp3}, Fig.~\ref{abgdepp1}, Fig.~\ref{bgdepp2}) agree with the analytic expressions \eqref{eq:res} i.e.
in the low temperature limit $x \rightarrow 1$ for class I,  $x \rightarrow (2-\zeta)/z$ for class II and $ x\rightarrow (2-\theta - \kappa \beta)/z$ for class III.
  For example, in Fig.~\ref{abgdepp3a}, the black, purple, and red curves correspond to class I, II and III respectively.   By using the values $(z, \theta,\zeta) = (\frac{21}{5}, \frac{7}{5}, -1)$ for the purple curve (class II) and  $(z,\theta,\zeta, \beta, \kappa) = {(\frac{49}{18}, \frac{2}{3}, -\frac{4}{3}, -\frac{2}{\sqrt{3}} , \frac{10}{3\sqrt{3}})}$ for the red curve (class III) we find that $x\sim 0.71$ and $x\sim 1.31$ respectively.\footnote{Here, we used \eqref{classIexp}, \eqref{classIIexp}, \eqref{classIIIexp} to compute $(z,\theta,\zeta, \beta, \kappa)$ from ($\alpha, \beta, \gamma$).} These agree with our numerical results. One may think that $x \sim 0.71$ is fine but $x \sim 1.31$ looks quite different from the value in our numerics (the red curve in the limit of $T/\mu \rightarrow 0$). However, this is because the red curve belongs to class III. As we showed in Fig.~\ref{fig:classIIIandcon} in class III the second term of \eqref{d2sigma} is dominant but the first term's contribution is still not negligible at low temperature. This contamination by the first term is reflected on our numerics. If we go to extremely low $T/\mu$ we will find a good agreement, which we have checked.\footnote{The same argument should work for class I, but we do not see a similar discrepancy to class III. This is because,  in class I, the first term and second term have the same power $x$ in $\rho \sim T^x$.} 
\end{enumerate}

\paragraph{The third potential in \eqref{ZandJV2}  ($\alpha<0$ and $\theta > 0$)}

The reference point is the black dot in Fig.~\ref{fig:pthregionb}, which is 
\begin{equation}
(\alpha,\beta,\gamma) = (\alpha_3,\beta_3,\gamma_3) := \left(-\frac{1}{\sqrt{3}}, -\frac{2}{\sqrt{3}}, \sqrt{3}\right) \ \ \Leftrightarrow  \ \ \left(z, \theta, \zeta\right) = (3, 1, -1) \,.
\end{equation}
Fig.~\ref{fig:pthregion} shows the allowed region of ($\beta, \gamma$) for a given $\alpha$: $\alpha = 1.4\alpha_3$ for Fig.~\ref{fig:pthregiona}, $\alpha=\alpha_3$ for Fig.~\ref{fig:pthregionb}, and $\alpha = 0.6\alpha_3$ for Fig.~\ref{fig:pthregionc}.

For fixed $(\beta, \gamma) = (\beta_3, \gamma_3)$, the $\alpha$ decreases from the purple dot ($\alpha= 1.4\alpha_3$) 
to the black dot ($\alpha= \alpha_3$) and to the red dot ($\alpha= 0.6\alpha_3$). For these three points the change of $x$ in $\rho \sim T^x$ is shown in Fig.~\ref{abgdepp3a}. The value of $x$ increases as $\alpha$ increases at low temperature $T/\mu$ while it does not change at high temperature $T/\mu$.

For fixed $(\alpha, \gamma) = (\alpha_3, \gamma_3)$, the $\beta$ increases from the blue dot ($\beta= 1.4\beta_3$) 
to the black dot ($\beta= \beta_3$) and to the orange dot ($\beta= 0.6\beta_3$). For these three points the change of $x$ in $\rho \sim T^x$ is shown in Fig.~\ref{abgdepp3b}. The value of $x$ decreases as $\beta$ increases in general, but it does not  change much in the intermediate temperature range $1 < T/\mu < 3$. 

For fixed $(\alpha, \beta) = (\alpha_3, \beta_3)$, the $\gamma$ increases from the dark yellow dot ($\gamma= 0.6\gamma_3$) 
to the black dot ($\gamma= \gamma_3$) and to the green dot ($\gamma= 1.4\gamma_3$). For these three points the change of $x$ in $\rho \sim T^x$ is shown in Fig.~\ref{abgdepp3c}. The value of $x$ increases as $\gamma$ increases.


In brief, these can be summrized as follow.  The region above (below) the dashed line corresponds to $x>1$ ($x<1$) in the relation $\rho \sim T^x$  in low temperature limit.  Our results show that this tendency remain at finite temperature in general.

From Fig.~\ref{abgdepp3} one might wonder if we decrease $\alpha$ and increase $\gamma$ the shift-up effect and shift-down effect cancel each other and the linear-$T$ may be obtained again in another point different from the black dot. Indeed, it is true as shown in the purple curve in Fig.~\ref{shift}. By this way, we have found that there is some range of parameters yielding the linear-$T$ resistivity.  However, this does not work in the other way, i.e. first increase $\alpha$ and then decrease $\gamma$. See the red curve in Fig.~\ref{shift}.
The difference between two ways is the class that the ending point arrives at. The former belongs to class II and the latter belongs to class III. As we showed in Fig.~\ref{fig:classIIIandcon} the dominant mechanism for conductivity changes as temperature increases in class III, so it is not easy to keep the linear-$T$ resistivity across this change. 

Because of very complicated coupled dynamics at finite temperature, it is not easy to have some analytic understanding of our observation. However, we speculate that it may have something to do with the vanishing potential near IR, $V \sim e^{\alpha \phi}$, for $\alpha<0$. For $\alpha \geqslant0$ the potential will diverge or be constant.

\paragraph{The first potential in \eqref{ZandJV2}  ($\alpha>0$ and $\theta < 0$)}


The reference point is the black dot in Fig.~\ref{fig:nthregion2}, which is 
\begin{equation}
(\alpha,\beta,\gamma) = (\alpha_1,\beta_1,\gamma_1) := \left(\frac{1}{3\sqrt{3}}, -\frac{2}{3\sqrt{3}}, \frac{5}{3\sqrt{3}}\right) \ \ \Leftrightarrow  \ \ \left(z, \theta, \zeta\right) = (5, -1, -3) \,.
\end{equation}
Fig.~\ref{fig:pthregion} shows the allowed region of ($\beta, \gamma$) for a given $\alpha$: $\alpha = 0.6\alpha_1$ for Fig.~\ref{fig:pthregiona}, $\alpha=\alpha_1$ for Fig.~\ref{fig:pthregionb}, and $\alpha = 1.4\alpha_1$ for Fig.~\ref{fig:pthregionc}. 

Similarly to the previous case $\alpha<0 (\theta >0)$, we consider six points around the reference black dot. For fixed $(\beta, \gamma) = (\beta_1, \gamma_1)$, the $\alpha$ increases from the purple dot ($\alpha= 0.6\alpha_1$) 
to the black dot ($\alpha= \alpha_1$) and to the red dot ($\alpha= 1.4\alpha_1$). For fixed $(\alpha, \gamma) = (\alpha_1, \gamma_1)$, the $\beta$ increases from the blue dot ($\beta= 1.4\beta_1$) to the black dot ($\beta= \beta_1$) and to the orange dot ($\beta= 0.6\beta_1$). For fixed $(\alpha, \beta) = (\alpha_1, \beta_1)$, the $\gamma$ increases from the dark yellow dot ($\gamma= 0.6\gamma_1$) 
to the black dot ($\gamma= \gamma_1$) and to the green dot ($\gamma= 1.4\gamma_1$)


As $\alpha$ increases, $\beta$ decreases, or $\gamma$ increases the curves shift up at finite temperature. See Fig.~\ref{abgdepp1}. It is again consistent with the fact that the region above (below) the dashed line corresponds to $x>1$ ($x<1$), where $x$ is defined in the relation $\rho \sim T^x$  in low temperature limit.
		
\paragraph{The second potential in \eqref{ZandJV2}  ($\alpha=\theta = 0$)}

The reference point is the black dot in Fig.~\ref{fig:zthregion}, which is 
\begin{equation}
(\alpha,\beta,\gamma) = (\alpha_2,\beta_2,\gamma_2) :=  \left(0,-\frac{1}{\sqrt{3}},\frac{2}{\sqrt{3}}\right) \ \ \Leftrightarrow  \ \ \left(z, \theta, \zeta\right) = (4 ,0 ,-2 ) \,.
\end{equation}
In this case, there is no change in $\alpha$ since $\alpha=\theta = 0$.

For fixed $(\alpha, \gamma) = (0, \gamma_2)$, the $\beta$ increases from the blue dot ($\beta= 1.4\beta_2$) to the black dot ($\beta= \beta_2$) and to the orange dot ($\beta= 0.6\beta_2$). For fixed $(\alpha, \beta) = (0, \beta_2)$, the $\gamma$ increases from the dark yellow dot ($\gamma= 0.6\gamma_2$) 
to the black dot ($\gamma= \gamma_1$) and to the green dot ($\gamma= 1.4\gamma_1$). 
Similarly to other cases, as $\beta$ decreases or $\gamma$ increases the curves shift up at finite temperature. See Fig.~\ref{bgdepp2}.


\section{Conclusion}\label{sec:conclusion}


In this paper, we studied \kyr{the linear-$T$ resistivity up to finite temperature in} more general cases: axion-dilaton theories or the EMD-Axion models. In order to study resistivity from low to  high temperature, we start with the low temperature limit or IR limit. In this limit, resistivity can be analyzed by the scaling geometries supported by the asymptotic potential and couplings in IR \eqref{IRpot}
\begin{equation}\label{IRpotcon}
V(\phi) \sim V_0 e^{\alpha \phi}\,, \qquad J(\phi) \sim e^{\beta \phi}\,, \qquad Z(\phi) \sim e^{\gamma \phi}\,.
\end{equation}
They are characterized by three parameters $\alpha, \beta, \gamma$, in terms of which, the necessary conditions for the linear-$T$ resistivity has been well studied in~\cite{Gouteraux:2014hca} and summarized in \eqref{eq:res}. In addition, there are many constraints for $\alpha, \beta, \gamma$ coming from physical conditions.  For instance, the specific heat should be positive and our geometry should be stable under small perturbation \cite{Gouteraux:2014hca, Ahn:2017kvc}. The constraints are expressed in \eqref{con000}, \eqref{c2const}, \eqref{constc3}, \eqref{con00000yy}, and \eqref{thetad}. Considering all, we have explicitly identified a parameter region which yields the linear-$T$ resistivity in low temperature limit. This region is displayed as 
a two dimensional surface in three dimensional $\alpha, \beta, \gamma$ space. See Fig.~\ref{fig:wregion}. 

To study resistivity at finite temperature, not only in the limit of low temperature, we have UV-completed $V(\phi)$ as
\begin{equation}\label{ZandJV2con}
\begin{split}
		&V(\phi) = \left\{ \begin{array}{lll}
		\frac{2d}{\alpha^2}\sinh^2\left(\frac{\alpha \phi}{2}\right) + (d+1)d\,,                                                   \, & \text{for} \, \theta <0\,,\\[0.5ex]
		(d+1)d\,,                                                                                                                                                                                  \,  & \text{for} \, \theta = 0\,,\\[0,5ex]
		d\left(\frac{1}{\alpha^2}+2\left(d+1\right)\right) \sech(\alpha \phi)-d\left(\frac{1}{\alpha^2}+\left(d+1\right)\right)\sech^2\left(\alpha \phi \right), \, &\text{for} \, d>\theta>0 \,.
		\end{array} \right.
	\end{split}
\end{equation}
with the same $J(\phi)$ and $Z(\phi)$ in \eqref{IRpotcon}.
Contrary to the low temperature limit, no analytic solution is available so we need to resort to numerical analysis.

By fine-gridding the surface of the linear-$T$ resistivity in low temperature limit,  in Fig.~\ref{fig:wregion}, we have systematically searched the parameters yielding the linear-$T$ resistivity up to high temperature. We found that the point
\begin{equation} \label{uytghu}
(\alpha, \beta, \gamma) = \left(-\frac{1}{\sqrt{3}}, -\frac{2}{\sqrt{3}}, \sqrt{3}\right) \ \ \Leftrightarrow \  \  (z, \theta, \zeta) = (3, 1, -1) \,,
\end{equation}
and its small neighborhood give the linear-$T$ resistivity from low temperature to high temperature if momentum relaxation is strong. 

\kyr{We have found \eqref{uytghu} by systematic and hard work by brute force. However, unfortunately, we do not have a deep understanding on  the {\it precise} physical mechanism for \eqref{uytghu} yet. The values (and the size of its neighborhood) will be changed with different UV completions. Thus, our main point is not the values \eqref{uytghu} but some qualitative discoveries we have obtained, which will be summarized as follows. }
\begin{enumerate}
\item Large momentum relaxation is a necessary condition to obtain robust linear-$T$ resistivity from low temperature up to high temperature. See Fig.~\ref{fig:z3t1vm}.
\item Among two terms in the conductivity formula \eqref{sigma}, the pair-creation term (the first term) is responsible for the linear-$T$ resistivity. In terms of geometry, this is nothing but the horizon value of the coupling $Z(\phi)$ with the Maxwell term for $d=2$ as shown in \eqref{action} and \eqref{d2sigma}, i.e.
\begin{equation}
e^{\gamma \phi(r_h)}  = r_h^{\gamma \kappa}\,.
\end{equation}
 It is a very simple formula. However, it is not easy to have a good intuition, because $r_h$ is a functions of ($T,  \mu, k $) determined by complicated coupled dynamics of various fields. Only in low temperature limit, $r_h \sim T$ so things are simplified. 
\item  In class III the dominant mechanism for conductivity is switched from the second (dissipation) term to first (pair-creation) term in the conductivity formula, as temperature increases. See Fig.~\ref{fig:classandcon}. Thus, we may expect that it is not easy to have a universal property, the linear-$T$ resistivity, in class III because of this mechanism change. Indeed, the parameters we found for linear-$T$ resistivity belong to class I and II. 
\item In class I, since both deformations (axion and charge) are marginal, the condition $T \ll$ Max($\mu, k$) is enough for the geometry to be 
captured by the IR scaling geometry. Thus, by increasing momentum relaxation $k$, it is more possible to have a larger range of linear-$T$ resistivity than the other classes.\footnote{We thank Blaise Gout\'{e}raux for pointing this out.}~\footnote{\kyr{Not all parameters in class I do not show this behavior. It may be partly understood by the fact the precise value of $\mu$ (not an order of magnitude) depends on the whole geometry, so depends on UV completion. Thus, we cannot simply say $T/\mu > 1$ or  $T/\mu < 1$ without knowing UV completion. When we say ``high temperature'', we do not mean $T/\mu \gg 1$, we mean, for example, $T/\mu = 4$ or $T/\mu = 6$, so the numerical value of $O(1)$ number matters.  Thus, we cannot simply judge $T/\mu > 1$ or  $T/\mu < 1$ without having a specific UV completion. }}
\item We also discussed how the conductivity behavior changes as $\alpha, \beta, \gamma$ changes for every potential in section \ref{sec43}. We speculate that the third potential in \eqref{ZandJV2con} may be favorable for the robust $T$ dependence of resistivity because the potential $V \sim e^{\alpha \phi}$ vanishes near IR for $\alpha < 0$. For $\alpha \geqslant0$ the potential will diverge or be constant.
\end{enumerate}

\begin{figure}[]
\centering
     {\includegraphics[width=7cm]{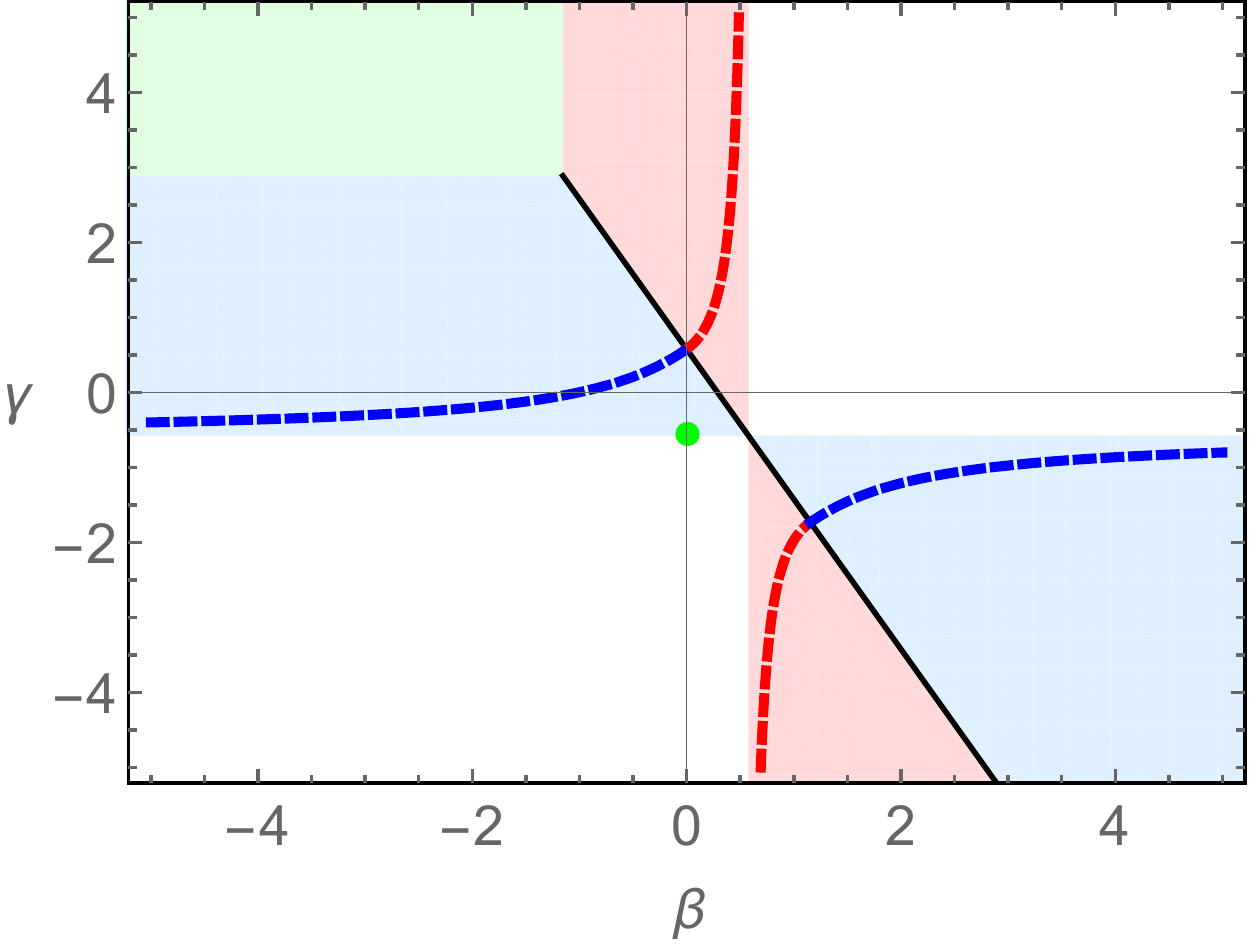} \label{adepp3}}
 \caption{The figure shows allowed ($\beta, \gamma$) region in the case of $\alpha = \frac{1}{\sqrt{3}}$. The black line represent class I case. The red, blue and green region stand for the class II, class III and class IV respectively. The dashed red(blue) line shows $(\alpha, \beta, \gamma)$ for the linear-$T$ resistivity in low temperature limit of class II(class III). The green dot $(\alpha, \beta, \gamma) = (\frac{1}{\sqrt{3}}, 0, - \frac{1}{\sqrt{3}})$.}\label{fig:Gub}
\end{figure}

It has been shown in~\cite{Jeong:2018tua} that the Gubser-Rocha model with linear axion exhibits the linear-$T$ resistivity up to high temperature. This model is also included in our set-up as shown in \eqref{GRwithA}.  It corresponds to the parameter set $(\alpha, \beta, \gamma) = (1/\sqrt{3}, 0, -1/\sqrt{3})$, which is the green dot in Fig.~\ref{fig:Gub}. 
It belongs to the first potential in \eqref{ZandJV2con} and class I.\footnote{It looks in class III because it is in the blue region. However, the boundary of the blue and red region is excluded in the definition of class II and class III. The green dot is a very special point, where $z \rightarrow \infty$, $\theta \rightarrow -\infty$ and $\eta := - \theta/z \rightarrow 1$. {The extension of the green dot to $\alpha$ direction corresponds to the case in section 3.2 in~\cite{Jeong:2018tua}.}} In this case, the DC conductivity ($\sigma_{DC}$) is
\begin{equation} \label{conduct1}
\sigma_{DC} =  \sqrt{1+\tQ} + \frac{\sqrt{1+\tQ}}{(k/\mu)^2} \,,
\end{equation}
where $\tQ$ is a complicated function of $(T,\mu,k)$.
Thus, the first (pair creation) and second (dissipation) term contribute in the same way through $\tQ$. However, it has been shown that the resistivity is linear in $T$ up to high temperature only for large momentum relaxation, $k/\mu \gg1$, so the first term dominates. This agrees with what we have found in this paper. 

In order to identify the power $x$ in $\rho \sim T^x$ in a more precise way we made a plot of   $\partial \ln \rho/\partial \ln T  $. This method is good enough to find a linear-$T$ behavior all the way from zero $T$ to high $T$. However, if there is a residual resistivity at zero $T$ (i.e. $T = \mathrm{constant} + T^x$) or if there is the linear-$T$ resistivity after some temperature $T_1$ (i.e. for $T > T_1 > 0$, $T = \mathrm{constant} + T^x$), our measure $\partial \ln \rho/\partial \ln T  $ may not be able to capture it. Thus, in such more relaxed conditions, which may be relevant for some phenomenology (for example, \cite{Cooper603}), there may be more parameter regime allowing the linear-$T$ resistivity in our model. 

Investigating conductivity at high temperature involves full bulk geometry so it depends on UV-completion of potential and couplings in general. In this paper, as a first step, we used a kind of minimal UV completion, in the sense that the potential $V$ depends on only one parameter $\alpha$.  As reviewed in Appendix \ref{appA} other UV completions are also possible. It will be interesting if we can find more conditions to constrain UV completion from other phenomenological input or theoretical consistency such as a top-down approach. Or, from phenomenological perspective, we may ask what kind of UV completion can allow the linear-$T$ resistivity up to high temperature.
	
	\acknowledgments
	We would like to thank Blaise Gout\'{e}raux, Yi Ling and Zhuo-Yu Xian for valuable discussions and comments. 
This work was supported by Basic Science Research Program through the National Research Foundation of Korea(NRF) funded by the Ministry of Science, ICT $\&$ Future Planning(NRF- 2017R1A2B4004810) and GIST Research Institute(GRI) grant funded by the GIST in 2019 and 2020.

\appendix
\section{UV completion}\label{appA}

In this appendix, we explain how to obtain the potential for UV completion in a little more detail. 
In principle, there are many ways to construct $V, Z$, and $J$ satisfying \eqref{IRpot} 
\begin{equation}\label{IRpotapp}
 J(\phi) \sim e^{\beta \phi}\,, \qquad Z(\phi) \sim e^{\gamma \phi}\,, \qquad V(\phi) \sim V_0 e^{\alpha \phi}\,,
\end{equation}
in IR for scaling geometry, and \eqref{cond10} and \eqref{cond11} 
\begin{align} 
	&V(\phi) = \frac{(d+1)d}{\ell_{AdS}^2} - \frac{1}{2}m^2 \phi^2 + \cdots, \label{cond10app} \\ 
	& Z(\phi) = 1 + \cdots  \,, \quad J(\phi) = 1 + \cdots, \label{cond11app}
\end{align} 
in UV for asymptotically AdS space.  In other words, for $V(\phi)$,
\begin{equation} \label{cond2app}
	V(0) = -2\Lambda =\frac{(d+1)d}{\ell_{AdS}^2}\,, \qquad V'(0) = 0\,, \qquad V''(0) = -m^2 = \frac{-\Delta(\Delta - d - 1)}{\ell_{AdS}^2} \,,
\end{equation}
where $\Delta$ is the conformal dimension of the dual operator of $\phi$. $\phi \rightarrow \infty$ in IR and $\phi \rightarrow 0$ in UV. 
%
%
%
%
 For simplicity, in this paper, we choose a minimal potentials studied in \cite{Ling2017}:
\begin{equation}\label{ZandJV1app}
Z(\phi) = e^{\gamma \phi}\,, \qquad J(\phi) = e^{\beta \phi}\,,
\end{equation}
and three cases for $V(\phi)$
\begin{equation}\label{ZandJV2app}
\begin{split}
		&V(\phi) = \left\{ \begin{array}{lll}
		\frac{2d}{\alpha^2}\sinh^2\left(\frac{\alpha \phi}{2}\right) + (d+1)d\,,                                                   \, & \text{for} \, \theta <0\,,\\[0.5ex]
		(d+1)d\,,                                                                                                                                                                                  \,  & \text{for} \, \theta = 0\,,\\[0,5ex]
		d\left(\frac{1}{\alpha^2}+2\left(d+1\right)\right) \sech(\alpha \phi)-d\left(\frac{1}{\alpha^2}+\left(d+1\right)\right)\sech^2\left(\alpha \phi \right), \, &\text{for} \, d>\theta>0 \,,
		\end{array} \right.
	\end{split}
\end{equation}
where  we set $\ell_{AdS} = 1$ for simplicity.

One way to determine the potential is to introduce several exponential terms as follows:
\begin{equation}
	V(\phi) = V_0 e^{\alpha \phi} + V_1 e^{\alpha_1 \phi} + V_2 e^{\alpha_2 \phi} + \cdots + V_n e^{\alpha_n \phi}\,.
\end{equation}
The simplest way is to use only first two terms, of which parameters $(V_0, V_1, \alpha_1)$ will be completely fixed by three conditions.  Instead, we may choose three  terms which is considered in \cite{Kiritsis:2015oxa}\footnote{{Ref.~\cite{Kiritsis:2015oxa} considers the 3 exponential terms rather than 2 exponential terms so that $\phi$ does not have a log behavior in  UV.}}:
\begin{equation}\label{3expapp}
	V(\phi) = V_0e^{\alpha \phi} + V_1 e^{\alpha_1 \phi}+V_2 e^{\alpha_2 \phi}\,.
\end{equation}
There are five free parameters $(V_0, V_1,\alpha_1, V_2, \alpha_2)$ and we have only three constraints. To fix two parameters we choose $\alpha_2=0$ and $V_0 = \frac{d}{2\alpha^2}$. There are two motivation for this choice: i) it avoids the log behavior of $\phi$ in UV~~\cite{Kiritsis:2015oxa}, ii) it can include the Gubser-Rocha model by choosing  $(\alpha, \beta, \gamma) = (1/\sqrt{3}, 0, -1/\sqrt{3})$ (see \eqref{GRwithA}). Furthermore, we also choose $m^2 = -d$
so the potential becomes only a function of $\alpha$. By these  choices, the potential $V(\phi)$ in \eqref{3expapp} becomes
\begin{equation}\label{3exp3app}
	V(\phi) = \frac{2d}{\alpha^2}\sinh^2\frac{\alpha \phi}{2}+(d+1)d \,.
\end{equation}
Because we want to have a form $V(\phi) \sim V_0 e^{\alpha\phi}$ in IR, \eqref{3exp3app} is valid for  $\alpha>0$ since $\phi >0$. Otherwise, the dominant term in IR will be $V(\phi) \sim V_0 e^{-\alpha \phi}$ or constant.

Although we may construct the potential by using 3 exponential terms \eqref{3expapp} for $\alpha < 0$, we use a different `building block' to construct the potential:
\begin{equation}
	V(\phi) = \frac{V_0}{2} \sech(\alpha \phi) + V_1 \sech^2(\alpha \phi) \,,
\end{equation}
where $1/2$ in $V_0/2$ is chosen to match the IR condition at $\alpha \phi \rightarrow -\infty$ \eqref{IRpotapp}. With this $\sech(\alpha \phi)$ form, one of the UV condition, the second of \eqref{cond2app} is automatically satisfied since $\phi \rightarrow 0$ in UV. Thus the remaining free parameters $V_0$ and $V_1$ are fixed by the first and third constraints in \eqref{cond2app}:
\begin{equation}
	V(\phi) = \left((d+1)d - \frac{V_0}{2}\right)\left(1-\tanh^2(\alpha \phi)\right) + \frac{V_0}{2 \cosh(\alpha \phi)} \,,
\end{equation}
where $V_0 = 2d(-{m^2}/{(d \alpha^2)} +2d + 2)$. Finally with   the choice of $m^2 = - d$, which is the same as the first  potential in Eq. \eqref{ZandJV2}, the potential becomes a function of $\alpha$ only. 

For $\alpha = 0$ case, we choose the constant potential:
\begin{equation}
	V(\phi) = (d+1)d\,,
\end{equation}
where $m = 0$. Even with this constant potential the field $\phi$ still runs nontrivially due to a non trivial $J$ and $Z$.

\bibliographystyle{JHEP}


\providecommand{\href}[2]{#2}\begingroup\raggedright\endgroup

\end{document}